%% file: ms.tex
\documentclass[prb,twocolumn,10pt,longbibliography, superscriptaddress]{revtex4}
\usepackage{amsmath}
\usepackage{graphicx}
\usepackage{color,soul}

\begin{document}
\input{./Cover}
\input{./Definitions}

\begin{abstract}
A single-electron transistor incorporated as part of a nanomechanical resonator represents an extreme limit of electron-phonon coupling.
While it allows for fast and sensitive electromechanical measurements, it also introduces backaction forces from electron tunnelling which randomly perturb the mechanical state.
Despite the stochastic nature of this backaction, under conditions of strong coupling it is predicted to create self-sustaining coherent mechanical oscillations.
Here, we verify this prediction using time-resolved measurements of a vibrating carbon nanotube transistor.
This electromechanical oscillator has intriguing similarities with a laser.
The single-electron transistor, pumped by an electrical bias, acts as a gain medium while the resonator acts as a phonon cavity.
Despite the unconventional operating principle, which does not involve stimulated emission, we confirm that the output is coherent, and demonstrate other laser behaviour including injection locking and frequency narrowing through feedback.
\end{abstract}
\maketitle


Backaction forces are an inescapable accompaniment to nanomechanical measurements.
While their ultimate limit is set by quantum uncertainty~\cite{Clerk2010}, in practical devices backaction often causes heating, damping, and dynamical instability even well before this limit is reached.
Among the most sensitive measurement probe for nanomechanics is the single-electron transistor (SET), which transduces motion with a precision that can approach the standard quantum limit~\cite{Schoelkopf1998a,LaHaye2004}.
However, the price is that the force exerted even by individual electrons modifies the mechanical dynamics.
This results in strong electron-phonon coupling~\cite{Mozyrsky2006,Steele2009,Lassagne2009}, leading to additional dissipation, frequency softening, nonlinearity, and cooling~\cite{Naik2006}.
Here, we show that the backaction force - due to stochastic single-electron tunnelling events - can also be harnessed to create a self-sustained oscillating state in a nanomechanical resonator.
The resulting device is analogous to a laser, where the optical field is replaced by the mechanical displacement.
In contrast to existing phonon lasers pumped by optical or mechanical drives~\cite{Vahala2009,Grudinin2010, Mahboob2013}, this oscillator is driven by a constant electrical bias.
The device exhibits several laser characteristics, detected via its electrical emission, including phase and amplitude coherence and injection locking.
The resulting oscillator serves both as a novel on-chip phonon source and to explore the connection between the physics of backaction and of lasers.

\begin{figure}[!ht]
\includegraphics[width=89mm]{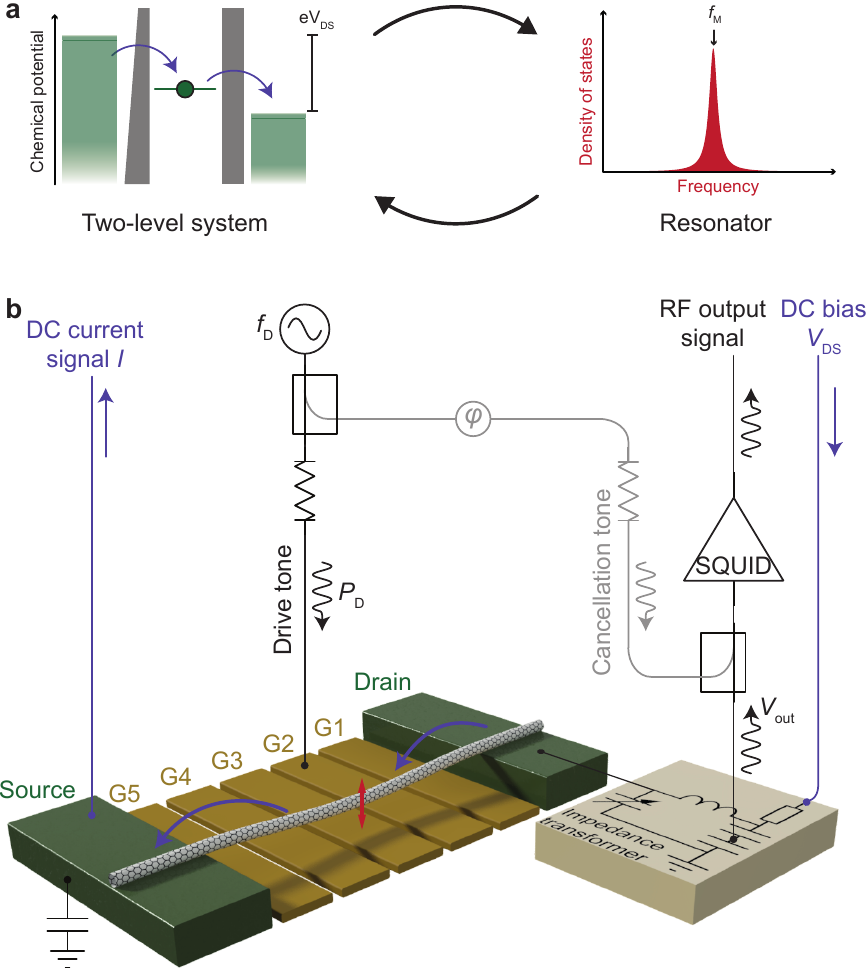}
\caption{\footnotesize{ {\bf Strongly coupled single-electron electromechanics.}
{\bf a}, Schematic of a single-electron transistor (SET) coupled to a mechanical resonator. The SET acts as a two-level system, while the resonator is a phonon cavity at mode frequency $\fM$. Electron tunnelling through the SET leads to a non-equilibrium population distribution which pumps the oscillator.
{\bf b}, Device realisation and measurement setup.
The vibrating nanotube, configured as an SET, is suspended between contact electrodes (green) and above gate electrodes (yellow).
The SET is biased by a drain-source voltage $V_\mathrm{DS}$, and the motion is measured via the electrical current, which is monitored both at DC (current path indicated by blue arrows) and via an RF circuit for time-resolved measurements (signal path marked by undulating arrows; see text and Supplementary Information).
The resonator can be driven directly by a tone with power $\PD$ at frequency $\fD$, part of which is routed via a cancellation path to avoid saturating the amplifiers.
}}
\label{Fig:1}
\vspace{-0.5cm}
\end{figure}

To enter this regime of strong backaction, the SET, serving as a two-level system, must couple strongly to a mechanical resonator serving as a phonon cavity (Fig.~\ref{Fig:1}a).
As a high-quality resonator, we use a suspended carbon nanotube~\cite{Sazonova2004}.
Nanotubes have both low mass and high mechanical compliance, which are favourable for strong electron-phonon interaction~\cite{Steele2009,Lassagne2009,Wen2018,DeBonis2018,Khivrich2019}.
The selected nanotube is a narrow-gap semiconductor, allowing the SET to be defined in the nanotube itself using tunnel barriers at each end and a conducting segment near the middle.
The two relevant SET states are the configurations with and without an excess electron.
Flexural vibration of the nanotube modulates the electrical potential experienced by the SET, causing the current to depend on the displacement; at the same time, each added electron exerts a force that is larger than both quantum and thermal force fluctuations (see Supplementary Information).

The combination of these effects sets up an electromechanical feedback with rich predicted behaviour~\cite{Armour2004}.
If the SET's energy splitting is resonant with the mechanical frequency, electrical excitations should be able to pump the resonator in a direct analogue of the micromaser~\cite{Rodrigues2007}.
More surprisingly, a laser-like instability is predicted even in a non-resonant situation, with complex dynamics that depend on level alignment and damping, and go beyond conventional laser behaviour~\cite{Bennett2006,Usmani2007}.
Previous experiments measuring time-average current through a nanotube have provided strong evidence for a threshold between resonance and oscillation~\cite{Huttel2009,Steele2009,Eichler2011}.
However, to test these predictions by fully characterizing the resulting states requires time-resolved displacement measurements~\cite{Tsioutsios2017, Barnard2019}, which have not yet been possible in this regime of strong backaction.
\\

\noindent
\textbf{Backaction turns a resonator into an oscillator}

To explore these dynamic effects, we implemented an electromechanical circuit for measuring the nanotube's vibrations directly~\cite{Wen2018} (Fig.~\ref{Fig:1}b).
The carbon nanotube is stamped across metallic contact electrodes to give a vibrating segment of length~$800$~nm~\cite{Wu2010}, and is measured at a temperature of 25~mK.
Voltages applied to five finger gates beneath the nanotube (labelled G1-G5) are used both to tune the electrical potential and to actuate vibrations by injecting an RF tone with drive power $\PD$.
A voltage bias $\VDS$ is applied between the contacts to drive a current $I$.
To configure the nanotube as an SET, the gate voltages are set to tune an electron tunnel barrier near each contact.
The conductance thus depends strongly on the displacement, which allows sensitive electromechanical readout via the current through the nanotube.
The radio-frequency (RF) part of the current is passed through an impedance transformer and then amplified, with the primary amplifier being an ultra-low-noise SQUID~\cite{Schupp2018}.
The resulting RF output voltage $\Vout$ is therefore proportional to the instantaneous displacement and is a sensitive time-resolved record of the mechanical vibrations~\cite{Wen2018}.

\begin{figure*}
\includegraphics[width=183mm]{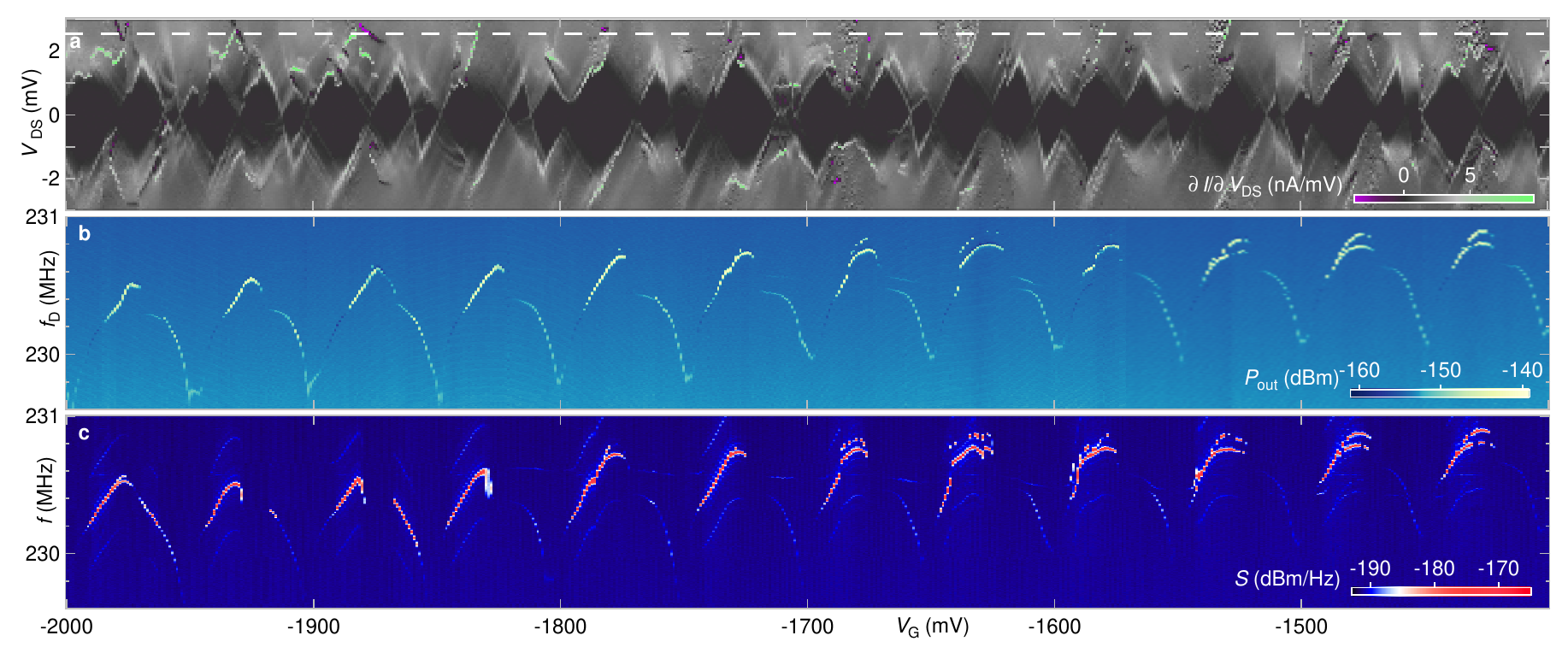}
\caption{\footnotesize{ {\bf Mechanical resonance and oscillation}
{\bf a}, Differential conductance of the nanotube as a function of gate voltage $\VG$ and bias $\VDS$, with no driving applied.
The diamond pattern  is characteristic of Coulomb blockade in an SET, with some irregularity due to electrostatic disorder.
Superimposed on the expected diamond pattern are sharp lines of strongly positive or negative differential conductance (green or purple in this colour scale), indicating thresholds for self-oscillation.
{\bf b}, Mechanical resonance detected in a transmission measurement.
The nanotube is biased with $\VDS=2.5$~mV (dashed line in {\bf a}) and driven with power $\PD = -99$~dBm to gate~G2.
The transmission is plotted as the emitted power $P_\mathrm{out}$ into the amplifier chain.
The mechanical resonance appears as a sharp spectral peak, or occasionally as a faint dip when RF leakage interferes destructively with the mechanical signal.
{\bf c}, Emission spectrum under the same conditions but with no RF drive.
A spectral peak is still present, at almost exactly the same frequency as in~{\bf b}.
This indicates self-driven mechanical oscillations.
Faint sidebands to the main signal are artefacts of the SQUID (see Supplementary Information).
}}
\label{Fig:2}
\vspace{-0.5cm}
\end{figure*}

To identify signatures of electromechanical feedback, we first measure the DC conductance as a function of bias and DC gate voltage $\VG$ applied to gate G2 (Fig.~\ref{Fig:2}a).
Superimposed on the diamond pattern characteristic of single-electron charging are irregular sharp ridges of strongly positive or negative conductance as the nanotube switches between high and low-current states.
Such features are associated with the onset of mechanical instability for bias exceeding a critical threshold~\cite{Steele2009, Usmani2007}.

We detect the mechanical resonance by fixing the bias voltage and measuring the transmission of the drive tone to the RF amplifier input.
When the drive frequency matches the mechanical resonance, the resulting motion relative to the gate electrodes changes the chemical potential of the SET, modulating the current at the drive frequency.
This current, entering the impedance transformer, excites an RF output voltage $V_\mathrm{out}$, which is proportional to the nanotube's displacement~\cite{Wen2018}.
The mechanical resonance therefore appears as a sharp peak in the electrical transmission from the drive to the output (Fig.~\ref{Fig:2}b).
The resonance frequency fluctuates quasiperiodically with gate voltage, which is a further indication of electromechanical coupling because the effective spring constant is softened when the SET is configured close to a Coulomb charge transition~\cite{Steele2009, Lassagne2009}.
From the peak width, the mechanical quality factor is $\QM \approx 2.1 \times 10^3$, with some gate voltage dependence because of electromechanical damping.

Mechanical oscillations, as distinct from a mechanical resonance, become evident when the output power spectrum is measured without driving (Fig.~\ref{Fig:2}c).
This undriven emission, plotted as a power spectral density~$S$ referenced to the amplifier input, shows a peak whose frequency approximately follows the resonance of Fig.~\ref{Fig:2}b.
The peak is only present for some gate voltage settings, and is brightest close to the transport ridges of Fig.~\ref{Fig:2}a.
Furthermore, this peak strengthens with increasing bias (see Supplementary Information).
For some gate voltage settings on the right of the graph, the peak switches between two or more frequencies, suggestive of dynamical bifurcation.
All these observations imply that the observed emission is a result of self-excited mechanical oscillations driven by the DC bias across the device.\\

\begin{figure}[hptb]
\includegraphics[width=89mm]{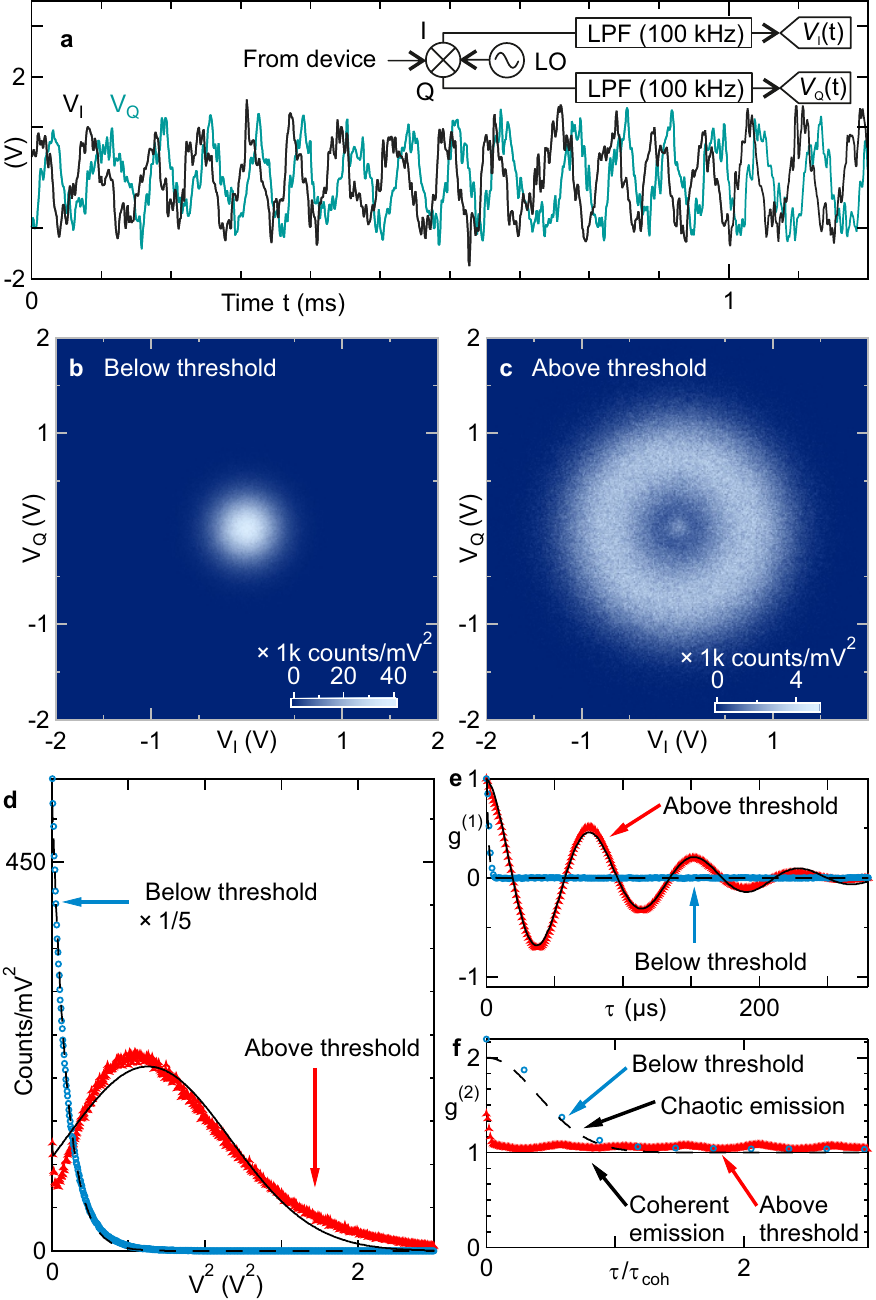}
\caption{\footnotesize{{\bf Coherence of the free-running oscillator. }
{\bf a},~Time traces of in-phase (I) and quadrature (Q) components demodulated from the oscillator output.
The heterodyne demodulation circuit is shown in the inset.
(LO: local oscillator; LPF: low-pass filter).
{\bf b},~Joint histogram of demodulated components with the gate voltage set below the oscillation threshold ($\VG=-1982$~mV).
{\bf c}, Histogram when configured above threshold ($\VG=-1568$~mV), showing the characteristic ring of coherent emission.
{\bf d}, Symbols: Histograms of total power $V^2(t) = \VI^2(t) + \VQ^2(t)$ below and above threshold, corresponding to the joint histograms in {\bf b, c}.
The former is scaled downwards for clarity.
Dashed line: fit to below-threshold data, assuming quasi-thermal source.
Solid lines: fits to above-threshold data.
Grey line assumes a Gaussian distribution of phonon numbers plus a small quasi-thermal fraction.
{\bf e}, Symbols: Autocorrelation as a function of time difference $\tau$.
Solid line: Above-threshold fit of the form $\gvv(\tau)=e^{-\tau/\tauc} \cos (2\pi \deltaf \tau)$, giving coherence time $\tauc=99~\mu$s and heterodyne frequency detuning $\deltaf=13$~kHz.
Dashed line: Similar fit to below-threshold data with $\deltaf$ fixed at zero, giving a decay time $\tauc= 3.3~\mu$s consistent with the filter bandwidth.
{\bf f}, Symbols: Second-order correlation, plotted with respect to the coherence time fitted above. Curves: Parameter-free predictions for Gaussian chaotic emission and for coherent emission~\cite{FoxBook}.
}}
\label{Fig:3}
\end{figure}

\noindent
\textbf{Mechanical coherence}

With fast electromechanical readout, the coherence of this mechanical oscillator can be directly confirmed by measuring the output signal in real time.
To do this, the signal is mixed with a local oscillator in a heterodyne circuit~\cite{Liu2015Science, Cassidy} to generate records of the in-phase and quadrature  voltages $\VI(t)$ and $\VQ(t)$ as a function of time~$t$.
The output record (Fig.~\ref{Fig:3}a) shows clear sinusoidal oscillations.
The onset of mechanical coherence is seen when the in-phase and quadrature time traces are represented as two-dimensional histograms for gate voltage settings above and below the oscillation threshold.
Below threshold, the histogram is peaked near the origin, consistent with a band-limited but quasi-thermal source such as a randomly kicked resonator (Fig.~\ref{Fig:3}b).
However, above threshold the histogram has a ring shape, showing amplitude coherence characteristic of a laser-like oscillator (Fig.~\ref{Fig:3}c).
The ring diameter corresponds to an approximate phonon number $\nphonon \sim 10^5$, i.e.\ an oscillation amplitude of $\sim 0.4$~nm, although there is a large uncertainty because of unknown device parameters (see Supplementary Information).

The clearest comparison to an ideal classically coherent source comes from a histogram of total output power, which is proportional to the number of phonons in the mode (Fig.~\ref{Fig:3}d).
Below threshold, the histogram follows the exponential distribution of completely incoherent quasi-thermal emission~\cite{Liu2015Science}.
Above threshold, the histogram shifts to a distribution where the most probable state has a non-zero output power, as expected for a coherent source.
It is approximately fitted by a Gaussian distribution, characteristic of a coherent oscillator in the limit of large phonon number $\nphonon$.
However, the distribution is slightly skewed, while its width, which for an ideal coherent state would be $\sim \sqrt{\nphonon}$, is much larger than expected.
Both the excess width and the skew indicate additional noise in the oscillator, presumably due to complex feedback between motion and electron tunnelling.
The faint spot at the centre of Fig.~\ref{Fig:3}c indicates bistability~\cite{Usmani2007,Pistolesi}, where the nanotube is either below threshold or has switched to a different frequency outside the measurement bandwidth.
The weight of the spot shows that for this gate setting the device spends approximately 0.5~\% of its time in such a state~\cite{Liu2015Science}.

While amplitude coherence is shown by the histogram, phase coherence is determined by plotting the autocorrelation of the demodulated signal $\gvv(\tau)$ as a function of time interval $\tau$ (Fig.~\ref{Fig:3}e).
For these settings, the data are well fitted by an decaying sinusoid, reflecting the slow phase drift of the free-running oscillator.
The envelope decay gives a phase coherence time $\tauc = 99~\mu$s, i.e.\ a coherence linewidth of $\delta f_\mathrm{coh} = 1/\pi \tauc = 3$~kHz, approximately three times narrower than the resonance linewidth $\fM/\QM$.
Coherence is further confirmed by plotting the second-order correlation function $g^{(2)}(\tau)$, which shows chaotic quasi-thermal behaviour below threshold but nearly coherent behaviour above threshold~\cite{FoxBook} (Fig.~\ref{Fig:3}f).

\begin{figure}[hptb]
\includegraphics[width=89mm]{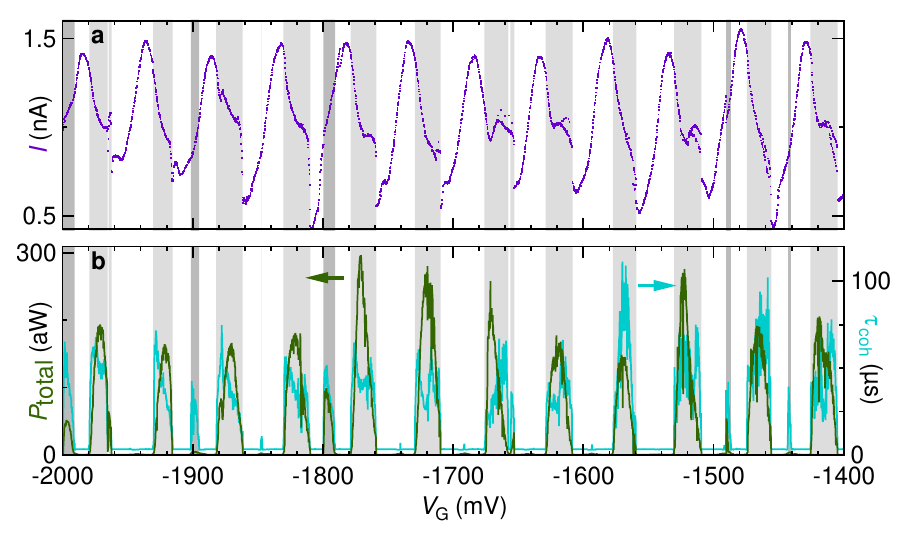}
\vspace{-0.5cm}
\caption{\footnotesize{{\bf Tuning the coherence with a gate voltage.}
{\bf a}, The DC current through the device as a function of gate voltage, with $\VDS = 2.5$ mV and no RF drive.
{\bf b}, Simultaneously acquired emission power $\Ptot$ (left axis) and fitted coherence time $\tauc$ (right axis).
Shading marks voltage settings with detectable emission.
}}
\label{Fig:4}
\end{figure}

As the gate voltage is swept, the device switches between oscillating and non-oscillating states, and both the power and coherence time change (Fig.~\ref{Fig:4}).
By simultaneously measuring the RF and DC signals, the consequences for DC transport can be seen.
Figure~\ref{Fig:4}a shows current as a function of gate voltage over several periods of Coulomb blockade, while Fig.~\ref{Fig:4}b shows the coherence time and emission power $\Ptot$ over the same range.
The oscillator switches on and off approximately once per Coulomb period.
Both the coherence time and the emitted power vary irregularly, but as expected most switches between oscillating and  non-oscillating conditions coincide with abrupt current changes.\\

\begin{figure*}[ht]
\includegraphics[width=183mm]{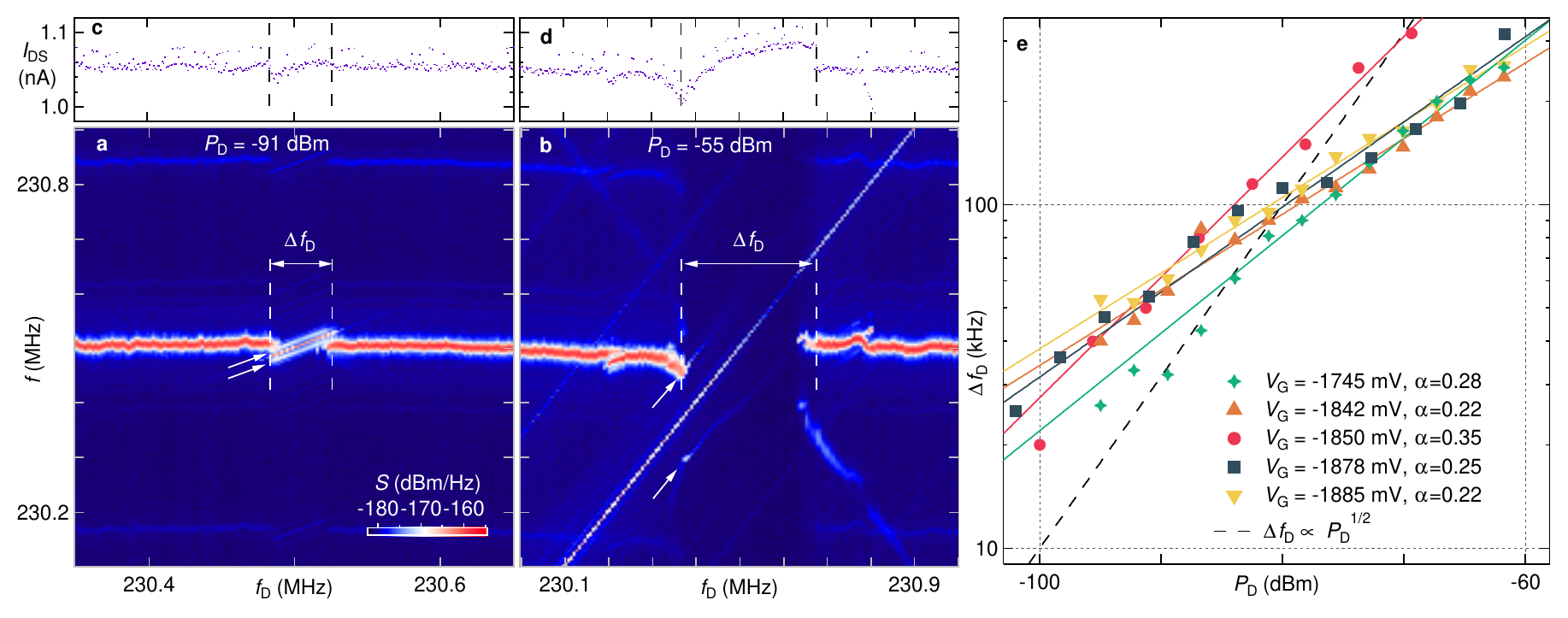}
\caption{\footnotesize{{\bf Injection locking of the nanomechanical oscillator.}
{\bf a,\,b}, Oscillator emission, plotted as a spectral density $S(f)$, in the presence of an injection tone at frequency $\fD$.
The broad horizontal line is the free-running emission.
When the injection frequency is within the capture range $\deltafD$, the oscillator locks to it, resulting in both a shift and a narrowing of the emission peak.
These two plots, measured with different injection power $\PD$, show that the capture range increases with increasing power.
A pair of unexpected power-dependent sidebands, discussed further in the Supplementary Information, is marked by arrows.
(Other faint sidebands running parallel to the main signal are artefacts of pickup in the SQUID.)
In panel {\bf b}, a distortion sideband is also evident running from upper left to lower right.
{\bf c,\,d}, DC current as a function of $\fD$, measured simultaneously with {\bf a} and {\bf b}.
{\bf e}, Locking range $\deltafD$ as a function of injection power $\PD$, measured for different gate voltage settings.
Symbols: Data;
Solid lines: Fits of the form $\fD \propto \PD^\alpha$, with $\alpha$ as a free parameter.
Dashed line: Dependence for $\alpha=1/2$, as expected for conventional injection locking~\cite{Adler1946}.
}}
\label{Fig:5}
\end{figure*}

\begin{figure*}[ht]
\includegraphics[width=183mm]{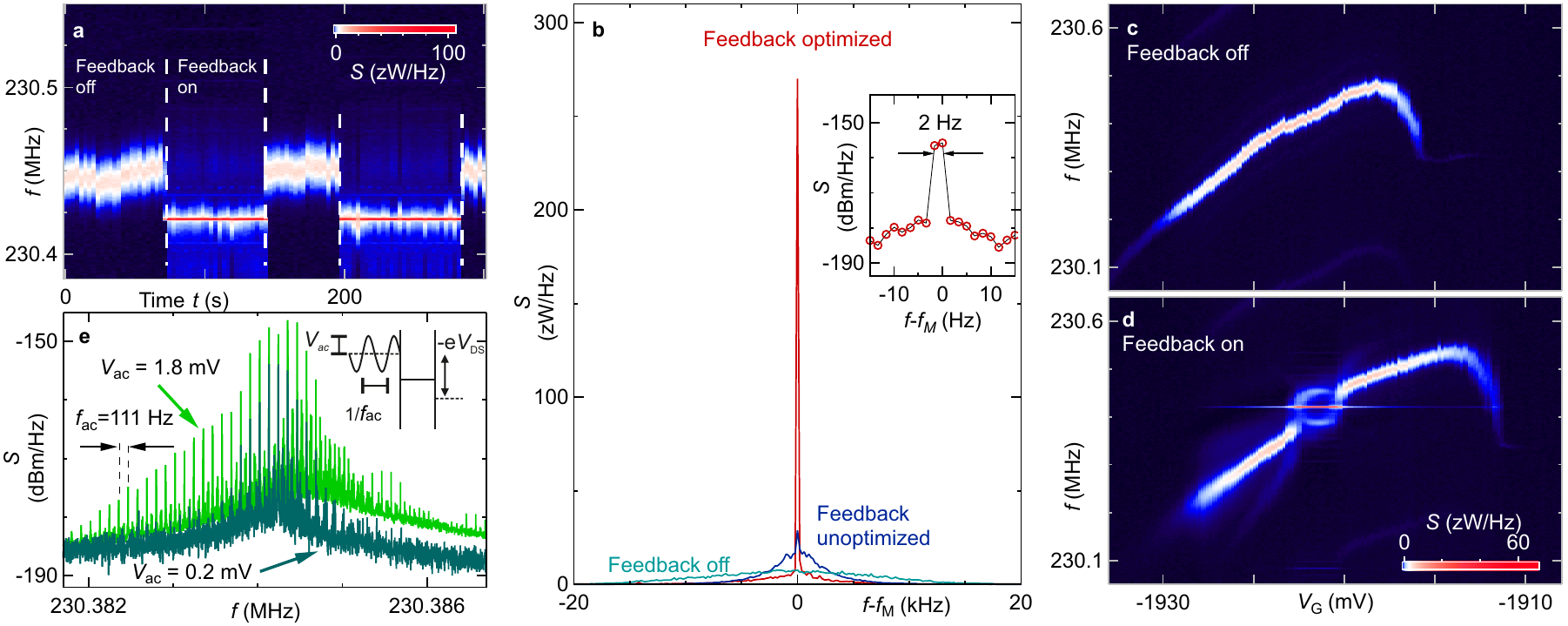}
\caption{\footnotesize{{\bf Stabilizing the oscillator with feedback}
{\bf a}, Power spectrum of oscillator emission $S(f)$ as a function of time.
With feedback off, the emission peak is broad and fluctuates.
Turning feedback on leads to intense emission at the target frequency, here 230.42 MHz.
To make the stabilised peak visible, the data is binned over 1~kHz vertical window in panels {\bf a,\,c}, and {\bf d}.
{\bf b}, Power spectrum for three different setting of the feedback controller's PID settings: feedback off, on but unoptimized, and fully optimized.
When optimally locked the linewidth is less than $2$ Hz (inset), limited by the point spacing.
{\bf c,\,d} Free-running ({\bf c}) and feedback-locked ({\bf d}) oscillation spectra as functions of DC gate voltage, showing the locking range.
With feedback on, a weak stabilised peak persists even when the oscillator's central frequency has moved outside the main locking range, indicating that occasional large frequency excursions occur that can be detected when they are temporarily stabilised by the feedback circuit.
{\bf e}
Using the stabilised oscillator as the basis for a microwave frequency comb.
Modulating the voltage bias (see inset) leads to a series of narrow, equally spaced sidebands, whose strength depends on the modulation depth.
In this case the comb extends over a range up to 35~kHz.
}}
\label{Fig:6}
\end{figure*}

\noindent
\textbf{Stabilised oscillations}

While the phase coherence time extracted from the autocorrelation characterizes the long-term oscillator stability, it is limited by slowly varying extrinsic effects such as charge noise or adsorbed atoms~\cite{DeBonis2018}.
To evaluate sensing schemes that rely on detecting mechanical frequency shifts, it important to identify the oscillator's intrinsic linewidth if this slow variation could be eliminated, which may be much narrower.
To measure the intrinsic linewidth, we employ two techniques from laser spectroscopy to stabilise the oscillator frequency.

First, we demonstrate that the oscillator can be locked to a stable but weak seed tone applied to the gate~\cite{Stover1966, Liu2015PRA}.
This phenomenon of injection locking, previously demonstrated for trapped ions~\cite{Knunz2010} and driven mechanical resonators~\cite{Seitner2017}, arises because feedback amplifies small forces close to the operation frequency.
In this measurement, the emission is monitored while the seed tone is applied at a nearby frequency $\fD$ (Fig.~\ref{Fig:5}).
As seen in Fig.~\ref{Fig:5}a,\,b, for a range of $\fD$ settings near the free-running oscillator's frequency and with sufficient drive power $\PD$, the broad emission line collapses onto the injection frequency.
The locking events are accompanied by steps in the DC current (Fig.~\ref{Fig:5}c, d).

The frequency range $\deltafD$ over which the oscillator is locked extends over many linewidths.
Figure ~\ref{Fig:5}e shows the locking range as a function of injected power, confirming that a stronger injection tone has greater frequency pull.
The data are well fitted by a power law of the form $\fD = A\PD^\alpha$, where $A$ and $\alpha$ are fit parameters.
However, whereas the theory of conventional oscillators~\cite{Adler1946} predicts an exponent $\alpha=0.5$, the data show a smaller exponent $\alpha \sim 0.3$, varying slightly with gate voltage but repeatable over two cooldowns.
Another unexpected observation is a pair of spectral sidebands, whose frequency offset, surprisingly, depends on injection power (see Supplementary Information).
These unexplained behaviours may be consequences of the stochastic nature of the current.

While injection locking clearly stabilises the oscillator's state, it also contaminates the output spectrum with the high-frequency seed tone.
An improved way to measure the oscillator's intrinsic linewidth is to use feedback to cancel out slow frequency wander.
This exploits the voltage tuning of the oscillator, and is analogous to Pound-Drever-Hall locking, used for sensitive laser measurements such as gravitational wave astronomy~\cite{Drever1983, Abbott2009}.
To implement this scheme (Fig.~\ref{Fig:6}), the oscillator is incorporated into a phase-locked loop using error signal voltage fed to gate G1 (see Methods).
Figure~\ref{Fig:6}a shows dramatic frequency narrowing when the feedback is turned on.
With optimised control parameters, the stabilised linewidth is $\delta f < 2$~Hz (Fig.~\ref{Fig:6}b), implying over $10^8$ coherent oscillations at the operating frequency of 230~MHz.
This represents an upper limit on the intrinsic linewidth when slowly varying environmental perturbations are cancelled, and is limited by the spectral resolution.
Similar to the Schawlow-Townes limit on a laser's linewidth~\cite{Schawlow1958}, the ultimate linewidth for an oscillator without stimulated emission is~\cite{Wiseman1999} $\delta f_\mathrm{ult} = \fM/4\nphonon Q_\mathrm{M} \sim 0.03$~Hz.

As expected, the feedback circuit succeeds in concentrating nearly the entire output into a narrow spectral line, provided that the oscillator's free-running frequency is close to the target frequency (Fig.~\ref{Fig:6}c,\,d)
The stabilisation range is set by the maximum feedback voltage.
However, feedback stabilises part of the emission even when this condition is not met, as seen by a weak spectral peak persisting beyond the expected voltage range~(Fig.~\ref{Fig:6}d).
This indicates that the oscillator occasionally deviates by several linewidths from its central frequency.
Feedback makes these excursions visible by temporarily capturing them.

Finally, we show how to generate a more complex output spectrum by exploiting the non-linear conductance of the SET.
A time-varying drain-source voltage modulates the amplitude of the oscillator's output, leading to regular sidebands spaced by the modulation frequency (Fig.~\ref{Fig:6}e).
Radio-frequency combs have been proposed for precise frequency comparison, as already used in optics~\cite{Cundiff2003}, and this nanomechanical comb is an alternative to devices based on superconducting resonators~\cite{Erickson2014} or Josephson junctions~\cite{Solinas2015, Cassidy}.\\

\noindent
{\bf Conclusion}

\noindent
The dynamical instability explored here is an extreme consequence of invasive displacement measurement.
For many kinds of nanomechanical sensing, it is a nuisance, because it means that same large bias necessary for precise measurement also strongly perturbs the displacement.
However, when the aim is to detect a small frequency shift (e.g.\ for mass spectrometry~\cite{Chaste2012} or some  force-detected magnetic resonance schemes~\cite{Stipe2001}), introducing feedback directly into the sensing element can be beneficial.
Clearly, the external frequency stabilization schemes described in the previous section are not directly useful for sensing  because they render the oscillator insensitive both to the undesirable drift and to the desirable signal (unless they can be separated spectrally).
However, even without applying external stabilization, the oscillation linewidth is narrower than the resonance linewidth, just as a laser's emission is narrower than its cavity linewidth~\cite{Bennett2006}, making small shifts easier to detect.

The similarities between SET nanomechanics and laser physics are intriguing~\cite{Bennett2006, Rodrigues2007}.
Like a laser, this device combines a pumped two-level system with a boson cavity, and shows phase and amplitude coherence as well as self-amplification.
It differs from a conventional laser by not requiring degeneracy between the SET and the resonator, since there is no stimulated emission.
A true phonon laser should emit directionally into a propagating sound wave~\cite{Maryam2013}, which this experiment (like previous phonon laser realisations~\cite{Vahala2009,Grudinin2010,Mahboob2013}) does not test.
However, to the extent that the key laser characteristic is output coherence~\cite{Wiseman1997}, this experiment does indeed realise a phonon laser.
It resembles unconventional lasers such as atom lasers that have coherent output statistics without stimulated emission~\cite{Ottl2005}.

Further development from this device could replace the SET with a coherent two-level system such as a double quantum dot~\cite{Brandes2003}, a superconducting SET~\cite{Rodrigues2007}, or an electron spin~\cite{Ohm2012, Palyi2012}.
This would allow a phonon laser driven by conventional stimulated emission.
Ultimately, superpositions might be transferred between the two-level system and the oscillator, allowing dynamic backaction to be studied in the fully quantum limit.

\vfill

\bibliography{./ref}

\vskip 0.25in
\noindent
{\bf Acknowledgements}
We acknowledge A. Bachtold, E. M. Gauger, Y. Pashkin, A. Romito and M. Woolley for discussions and T. Orton for technical support.
This work was supported by EPSRC (EP/N014995/1, EP/R029229/1), DSTL,  Templeton World Charity Foundation, the Royal Academy of Engineering, and the European Research Council (grant agreement 818751).
\\ \\
{\bf Methods}

{\bf IQ tomography}. To generate the time traces in Fig.~\ref{Fig:3}a, the amplified emission is mixed with a local oscillator running at a nominal frequency offset by $\deltaf= 15$~kHz below the mechanical frequency.
The local oscillator and the IQ mixer are implemented in a Zurich Instruments UHFLI lock-in amplifier.
The mixer's two intermediate-frequency outputs, corresponding to the two quadratures of the signal and oscillating at frequency $\deltaf$, are low-pass filtered with a 100~kHz cutoff to generate the time traces of Fig.~\ref{Fig:3}a. Histograms and autocorrelation traces are built up from 1\,s of data at each voltage setting. For the data of Fig.~\ref{Fig:4}, in which the oscillator frequency changes with gate voltage, the local oscillator is adjusted at each voltage setting to maintain approximately constant frequency offset $\Delta f$.

The histograms resulting from these time traces (Fig.~\ref{Fig:3}d) were fitted as follows~\cite{Liu2015Science}, using the fact that $V^2(t)$ is proportional to the number of phonons. Below threshold, the histogram was fitted assuming a quasi-thermal distribution
\begin{equation}
P(V^2) \propto e^{-V^2/\VT^2},
\end{equation}
with $\VT$ as a free parameter.
Above threshold, the fit is
\begin{equation}
P(V^2) \propto e^{-(V^2-V_0^2)^2/\sigma_V^4} + A_\mathrm{T} e^{-V^2/\VT^2}
\end{equation}
with $V_0$, $\sigma_V$, $\VT$ and $A_\mathrm{T}$ as free parameters, where the two terms represent a Gaussian distribution over phonon numbers and a small thermal contribution, respectively.

{\bf Signal autocorrelation}.
The autocorrelation is defined as
\begin{equation}
\gvv(\tau) \equiv \frac{\langle \VI(t) \VI(t+\tau)\rangle}{\langle \VI^2(t) \rangle},
\end{equation}
where the expectation value is calculated over a long time trace.
In Fig.~\ref{Fig:3}e, this function is fitted with the exponentially decaying oscillation stated in the caption, with $\tauc$ and $\Delta f$ as fit parameters.
While the fit here is good, at other gate voltages the oscillator sometimes jumps between different frequencies during data acquisition. For Fig.~\ref{Fig:4}b, a more general function is therefore used:
$
\gvv (\tau) =\mu e^{-\tau/\tauc} \cos (2\pi \deltaf \tau) + (1-\mu)e^{-\tau/\taufilter}.
$
The first term represents the contribution of the oscillator running at its primary frequency, and the second term represents contributions from other frequencies outside the detection bandwidth. The additional fit parameters are $\mu$, the fraction of time at the primary oscillation frequency, and $\taufilter$, the decay time of the other contributions.

The second-order correlation function~\cite{FoxBook} is
\begin{equation}
g^{(2)}(\tau) \equiv \frac{\langle V^2(t) V^2(t+\tau)\rangle}{\langle V^2(t) \rangle^2}.
\end{equation}
For a perfectly coherent source, $g^{(2)}(\tau)=1$ at all $\tau$, whereas Gaussian chaotic emission has $g^{(2)}(\tau)=1+e^{-\pi(\tau/\tauc)^2}$. These are the functions plotted in Fig.~\ref{Fig:3}f.

{\bf Feedback stabilization}. In the phase-locked loop used for Fig.~\ref{Fig:6}, the amplified emission is first mixed with a local oscillator running at the target frequency to generate a quadrature voltage proportional to the phase error.
This error signal is digitised at up to 14.06~MHz and used as input for a proportional-integral-derivative (PID) controller~\cite{Ogata1970} to generate a correction voltage.
The correction voltage is filtered with a 350~Hz low-pass cutoff and clipped to a range of $\pm 0.8$ mV, before being fed back to gate G1 of the device.

\end{document}


\input{./Cover}
\input{./Definitions}

\parbox[c]{\textwidth}{\protect \centering \Large \MakeUppercase{Supplementary Material}}
\rule{\textwidth}{1pt}
\vspace{1cm}

\date{\today}

	\setcounter{equation}{0}
	\setcounter{figure}{0}
	\setcounter{table}{0}
	\setcounter{page}{1}
	\makeatletter
	\renewcommand{\thesection}{S\Roman{section}}
	\renewcommand{\theequation}{S\arabic{equation}}
	\renewcommand{\thefigure}{S\arabic{figure}}
	\renewcommand{\thetable}{S\Roman{table}}
	\renewcommand{\bibnumfmt}[1]{[S#1]}
	\renewcommand{\citenumfont}[1]{S#1}

\newcommand{\qbar}{\overline{q}}
\newcommand{\wwidth}{183mm}

\maketitle

\section{Theory of self-driven oscillations}

\subsection{Oscillation mechanism \label{Sec:model}}

Here we describe a simple model of the self-oscillation mechanism, and explain why the gate voltage settings that lead to self-oscillations usually also coincide with local maxima of the resonance frequency (as seen in Fig.~2b,\,c of the main text).
The model works in the limit of a slow oscillator, where the phonon energy is much smaller than the energy splitting of the SET.
We take a simplified approach to self-oscillations mediated by Coulomb blockade, following especially Refs.~\cite{Bennett2006,Usmani2007}, which present much more complete theory.

\begin{figure}[ptb]
	\includegraphics{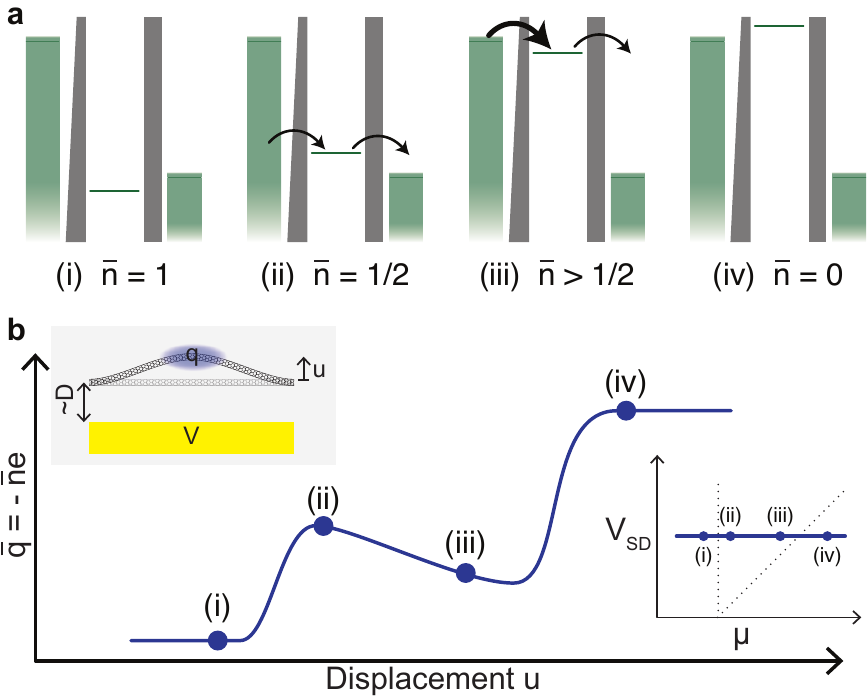}
	\caption{\footnotesize{\bf Mechanism of self-oscillations.}
		{\bf a}, Chemical potential of dot and leads for four different values of nanotube displacement (assuming gate voltage $V>0$, and considering only a single level within the bias window). Tunnel barriers are marked by vertical bars, with changing bar thickness denoting energy-dependent tunnel rates. The mean excess electron number $\overline{n}$ is shown for each displacement.
		{\bf b}, Quantum dot mean charge as a function of displacement. The four configurations sketched in {\bf a} are marked. Near configuration (iii), the condition $V\qbar'<0$ occurs, permitting self-oscillations. Lower inset: The same configurations plotted in a dot stability diagram as a function of bias and chemical potential relative to the right lead. Dotted lines mark the alignment of the dot potential with the leads, i.e.\ the edges of Coulomb blockade diamonds. Upper inset: Sketch of quantum dot displacement above gate.
		\label{FigS:OscilMech}
	}
\end{figure}

Consider a single quantum dot suspended above a gate held at positive potential~$V$, representing the average potential of the five gates in this experiment (Fig.~\ref{FigS:OscilMech}). The equation of motion for the displacement $u$ as a function of time $t$ is:
\begin{equation}
m\ddot{u}(t) = - m\OmegaZero^2 u(t) - \frac{m \OmegaZero}{Q_0} \dot{u}(t) + \frac{q(t)V}{D}
\label{eq:EOM}
\end{equation}
where $\dot{u}$ denotes a time derivative, $m$ is the suspended mass and $\OmegaZero=2\pi\fM$ and $Q_0$ are the bare mechanical resonance frequency and quality factor. The last term describes the electrostatic force on the suspended device: $q(t)$ is the instantaneous charge on the dot and $D$ is a distance (on the order of the separation between nanotube and gate) that relates the gate voltage~$V$ to the electric field experienced by the dot.

Electromechanical feedback occurs because the dot charge depends on the displacement~\cite{Blanter2004}. For a very open quantum dot, the characteristic electron tunnel rate $\Gamma$ is much larger than the mechanical frequency ($\Gamma \gg \OmegaZero$) and so the charge in Eq.~\eqref{eq:EOM} can be replaced by its instantaneous classical expectation value, set by the displacement:
\begin{equation}
q(t)\rightarrow \overline{q}(u(t)).
\end{equation}
However, if the tunnel rate approaches the mechanical frequency ($\Gamma \gtrsim \OmegaZero$), then we must take account of the time needed to reach an equilibrium charge on the quantum dot, given the changing displacement. The charge lags the displacement by approximately the electron tunnelling time. This is modelled by expanding $\overline{q}$ to lowest order in~$\Gamma/\Omega$:
\begin{align}
\overline{q}(u(t))	\rightarrow \overline{q}&(u(t-1/\Gamma))	\\
	\approx		 \overline{q}&(u(t) - \dot{u}(t)/\Gamma)
\end{align}
where $\Gamma$ sets the timescale for the charge on the dot to reach equilibrium. For small displacement, this can further be expanded in $u$, giving
\begin{equation}
\overline{q}(u(t)) \rightarrow \overline{q}(0) + \overline{q}'(0)[u(t)-\dot{u}(t)/\Gamma],
\end{equation}
where the prime denotes a spatial derivative.
Substituting into Eq.~\eqref{eq:EOM} then leads to a modified equation of motion, where
the electromechanical coupling modifies the resonance frequency and the quality factor as follows:
\begin{align}
\Omega 	\rightarrow&\; \OmegaZero + \Delta \Omega \label{eq:modifiedOmega}\\
Q 		\rightarrow&\; Q_0 (1-2Q_0\Delta \Omega/\Gamma)^{-1},
\label{eq:modifiedQ}
\end{align}
where
\begin{equation}
\Delta \Omega = -\frac{V}{2m\OmegaZero D}\,\qbar'(0).
\label{eq:deltaOmega}
\end{equation}
In the limit of instantaneous electron tunnelling ($\Gamma \rightarrow \infty$), the resonator's frequency is modified, but not its damping. However, for a slower tunnelling rate, Eq.~\eqref{eq:modifiedQ} shows that electron tunnelling also modifies the damping factor. Self-driven oscillations occur when the damping becomes negative.

The usual situation is for the quantum dot charge to increase monotonically as the chemical potential is decreased, i.e.\ $V\qbar'>0$. Equation~\eqref{eq:deltaOmega} shows that the mechanical resonance is then softened and the damping increased, especially near Coulomb peaks where $|\qbar'|$ is large~\cite{Steele2009, Lassagne2009}. However, at large source-drain bias ($|V_\mathrm{SD}| > k_\mathrm{B}T, \Gamma$), there may be gate voltage configurations where charging is reversed ($V\qbar'<0$), leading to an increase in mechanical frequency. Equation~\eqref{eq:modifiedQ} predicts that self-oscillations occur when
\begin{equation}
V\qbar' < -\frac{m\OmegaZero\Gamma D}{Q_0}.
\label{eq:OscillationCondition}
\end{equation}
This situation can arise if tunnel rates depend on energy~\cite{Usmani2007,Meerwaldt2012}. As shown in Fig.~\ref{FigS:OscilMech}(a), a change in barrier asymmetry with energy causes the quantum dot occupation to vary non-monotonically, allowing the condition Eq.~\eqref{eq:OscillationCondition} to occur~(Fig.~\ref{FigS:OscilMech}(b)).

This model predicts that the same gate voltage settings that allow self-oscillations will also lead to positive frequency shifts $\Delta \Omega$.
This is seen in Fig.~2{b,\,c}, which plots the driven and undriven power spectrum as a function of gate voltage.
The peaks in the undriven spectrum, which are the signature of self-oscillation, correspond to gate voltage ranges where the driven mechanical frequency is above its trendline, indicating that $\Delta \Omega$ is positive.
This is as predicted by Eq.~\eqref{eq:modifiedQ}, supporting this electromechanical mechanism as the main driver for the oscillations.

This simple model assumes just two tunnel barriers in the nanotube and therefore a single quantum dot, whereas in a real device disorder along the nanotube can lead to multiple dots.
In fact, Fig.~2 provides some evidence of this: the Coulomb diamonds in Fig.~2a are irregular, and the periodicity in Figs.~2b,\,c (measured at a bias larger than the charging energy) does not match the periodicity in Fig.~2a.
If so, this may be the reason for the energy-dependent tunnel rate, since tunnelling between two quantum dots depends on their energy alignment.
Thus, some electrostatic disorder may be necessary to satisfy the oscillation criterion Eq.~(\ref{eq:OscillationCondition}).

While this mechanism appears to explain all our data, it is possible that other oscillation mechanisms also contribute.
In other nanomechanical resonators, oscillations can emerge through a photothermal mechanism~\cite{Barton2012,Tavernarakis2018} and an analogous electrothermal effect might contribute here since the ohmic heating depends on displacement.
However, if this were the only mechanism then the feedback should only be positive for one sign of $\partial I/\partial u$ and therefore of $\partial I/\partial \VG$, whereas Fig.~3 of the main text shows that oscillations occur in regions of both positive and negative current slope.

\subsection{Quantifying the electromechanical coupling}
To compare this nanotube device with other nanomechanical resonators, in this section we quantify the electromechanical coupling.
The coupling arises from the extra force exerted on the nanotube by each additional electron on the SET.
Modelling the device for simplicity as a single quantum dot, this force is
\begin{equation}
\Fem = \frac{1}{\CS+\CNT+\CD}\frac{e}{V}\frac{\partial\CNT}{\partial u}
\end{equation}
where $\CS,\CNT,\CD$ are the capacitance between to source, gate, drain, respectively.
This force can be compared with the quantum zero-point force scale, defined by analogy with the zero-point displacement $\uZPM \equiv \sqrt{\hbar/2m\OmegaZero}$ as
 $\FZPM \equiv m\OmegaZero^2\,\uZPM=\sqrt{\hbar m \OmegaZero^3/2}$.
 This leads to a dimensionless electromechanical coupling parameter defined as~\cite{Usmani2007}:
\begin{align}
	\lambda &\equiv \frac{1}{2} \left(\frac{\Fem}{\FZPM}\right)^2= \frac{\Fem^2}{\hbar m \OmegaZero^3} \sim 40.
	\label{eq:lambda}
\end{align}
The uncertainty of this estimate is large because of unknown device parameters such as the nanotube mass and diameter. Based on previous characterization of the same device~\cite{Wen2018}, we find that the coupling parameter is in the range $0.6 < \lambda < 6000$.
Thus it is probable that the electromechanical force is larger than quantum fluctuations, i.e.\ the device is in the regime of strong coupling.
This coupling parameter $\lambda$ can also be interpreted as the ratio between the polaronic energy~\cite{Micchi2015} $\Epolaron\equiv\Fem^2/m\OmegaZero^2$ and the phonon energy $\hbar \Omega_0$.

The electromechanical force can also be compared with the thermal fluctuation force, given by $F_\text{th}=\sqrt{m\OmegaZero^2\kB T}$, where $\kB$ is Boltzmann's constant and $T$ is the temperature. Taking $T=25$~mK (i.e.\ assuming that the resonator is fully thermalised), we can calculate a corresponding thermal coupling parameter
\begin{equation}
\lambda_\text{th} \equiv \frac{1}{2} \left(\frac{\Fem}{F_\text{th}} \right)^2 = \frac{\hbar \OmegaZero}{2\kB T} \lambda \sim 9,
\end{equation}
showing that the electromechanical force fluctuations probably also dominate the thermal fluctuations.

The model of Sec.~\ref{Sec:model} is based on Refs.~\cite{Bennett2006,Usmani2007} which were developed for the weak-coupling regime $0<\lambda \ll 1$. Our experimental results show that these models give qualitatively accurate predictions even beyond that range, when $\lambda$ approaches or exceeds unity.

\subsection{Phonon number and minimal linewidth}

\subsubsection{Phonon number}
In this section we estimate the number of phonons in the oscillator.
Similar to a laser, the mean phonon number $\nphonon$ can be deduced from the emission power $\Ptot$ using
\begin{equation}
\nphonon = \frac{\Ptot}{\kappa h \fM},
\label{eq:npalgebraic}
\end{equation}
where $\kappa \equiv 2\pi \fM / \QM$ is the mechanical decay rate.
This equation assumes that all of the oscillator's energy is lost by electrical emission, and is therefore a lower bound on the phonon number.
From the linewidth in Fig.~2b, we extract
$\kappa = 2\pi\times (11\pm7)$~kHz.
The emitted power $\Ptot$ is extracted from a Lorentzian fit to the emission spectrum.
For Fig.~4c, this leads to a phonon number
\begin{equation}
	\nphonon = 0.9_{-0.3}^{+2}\times10^5,
	\label{eq:npnumeric}
\end{equation}
with the error bar dominated by the uncertainty in $\kappa$.
The corresponding oscillation amplitude can be calculated from the zero-point amplitude:
\begin{align}
	\uRMS = \uZPM\sqrt{\nphonon} = 400_{-200}^{+900}~\text{pm},
\end{align}
again subject to large uncertainty in the device parameters~\cite{Wen2018}.

\begin{figure*}[ptb]
\includegraphics[width=\wwidth]{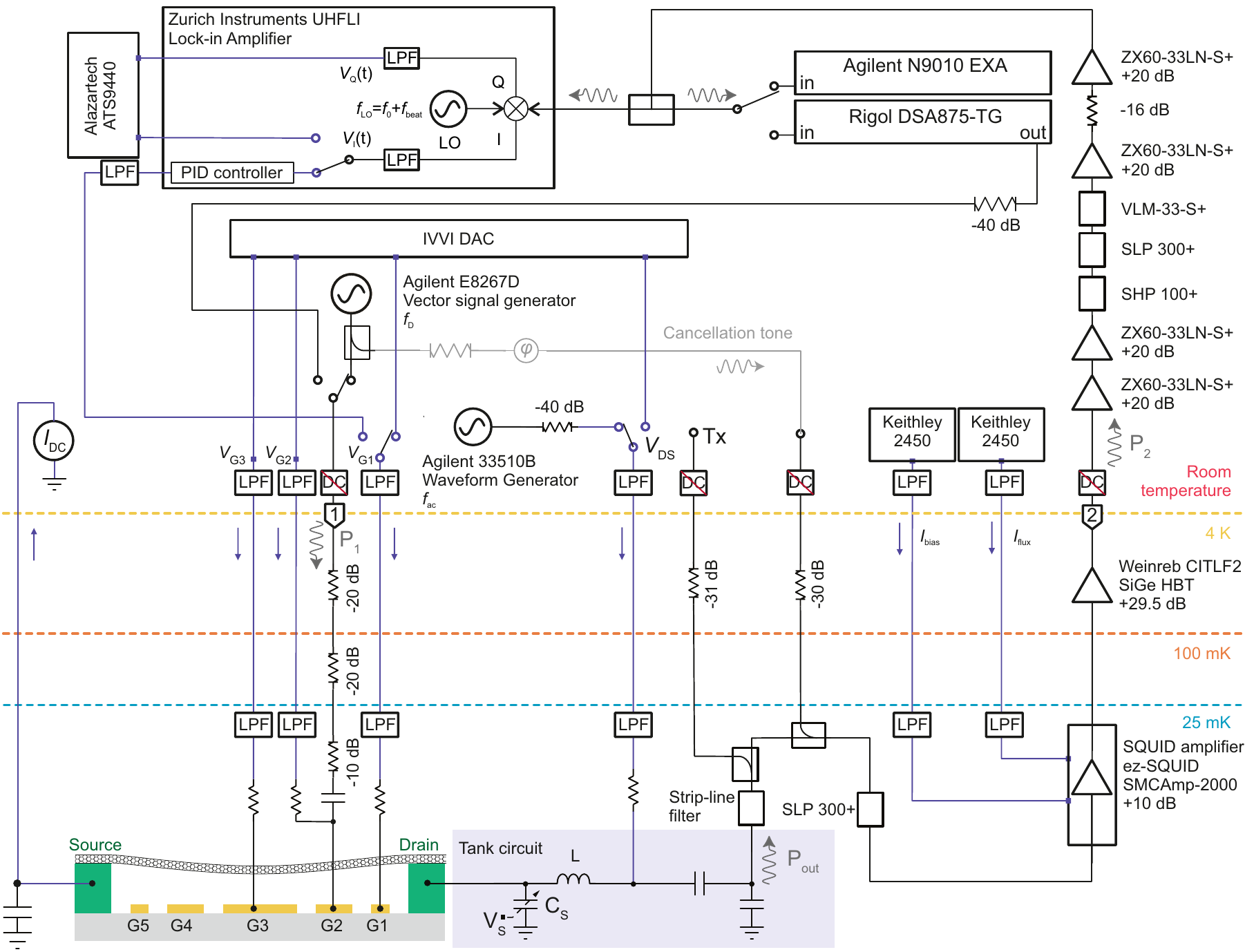}
\caption{\footnotesize{{\bf Full circuit diagram of the setup,} as explained in Section~\ref{Sec:setup}.
The abbreviation LPF denotes a low-pass filter.
 }}
\label{FigS:Circuit}
\vspace{-0.5cm}
\end{figure*}

\subsubsection{Minimal linewidth}
Quantum uncertainty predicts a minimal linewidth for an oscillator containing a given number of quanta.
For a conventional laser, this is the famous Schawlow-Townes linewidth~\cite{Schawlow1958}.
For a general oscillator, the limit is reduced by a factor of two, giving the ultimate linewidth
\begin{equation}
\delta f_\text{ult}=\frac{\kappa/2\pi}{4\nphonon}.
\label{eq:fult}
\end{equation}
This limit is set ultimately by the quantum statistics of the boson mode, and is therefore independent of the oscillator mechanism~\cite{Wiseman1999}.
For the configuration of Fig.~6, the emission power is similar to that in Fig.~4c, leading to a similar phonon number (Eqs~(\ref{eq:npalgebraic})-(\ref{eq:npnumeric})).
Equation (\ref{eq:fult}) therefore predicts $\delta f_\text{ult} \approx 0.03$~Hz.

\section{Experimental setup \label{Sec:setup}}

In Fig.~\ref{FigS:Circuit} we show the complete experimental setup.
The experiment uses a Triton 200 dilution refrigerator with a base temperature of $\sim 25$~mK.
The nanotube quantum dot is configured using DC voltages supplied from a Delft IVVI rack, and is operated probably in a hole-doped configuration.
For driven motion, an RF tone, applied to port~1 of the cryostat, passes via an attenuated coaxial line and a bias tee to gate G2.

The DC component of the  current through the nanotube is monitored by a current-to-voltage amplifier whose output is acquired by an analog-to-digital converter (ADC, Alazartech ATS9440).
The RF component of the current, which is proportional to the nanotube's instantaneous displacement, is transduced to a voltage using a tank circuit and then amplified using a SQUID amplifier at the cryostat base temperature followed by by a SiGe heterostructure bipolar transistor amplifier at 4~K.
The SQUID is current- and flux- biased using a pair of Keithley 2450 sources.
The RF signal, emerging from port 2 of the cryostat, is further amplified and filtered at room temperature before being passed to an RF lock-in amplifier (Zurich Instruments UHLFI) and to the input of a spectrum analyser (Agilent N9010EXA) or scalar network analyser (Rigol DSA875-TG)
This RF electromechanical readout is described in Ref.~\cite{Wen2018}, and the SQUID is described in Ref.~\cite{Schupp2018}.

The transmission measurements of Fig.~2b are performed using the scalar network analyser to inject a tone into port 1 and measure the resulting amplified signal.
For self-oscillation measurements (Figs 2c, 5, and 6), the signal is acquired using the spectrum analyser.
For the coherence measurements of Figs. 3-4, the signal is demodulated by the lock-in amplifier to yield in-phase and quadrature voltages, which are passed to the ADC.
For injection locking (Fig.~5), the tone is injected at port 1.
In this measurement, a phase-shifted cancellation tone is added to the SQUID input to prevent amplifier saturation.
For feedback stabilisation (Fig.~6), a PID controller integrated into the lock-in amplifier is used to apply a feedback voltage (low-pass filtered to 350~Hz) to gate G1.
For the frequency-comb experiment of Fig.~6e, the oscillating drain-source bias was applied using a waveform generator (Agilent 33510B).

\section{Supplementary data}

\subsection{Effect of bias on oscillation amplitude}
\begin{figure*}[b!]
\includegraphics[width=183mm]{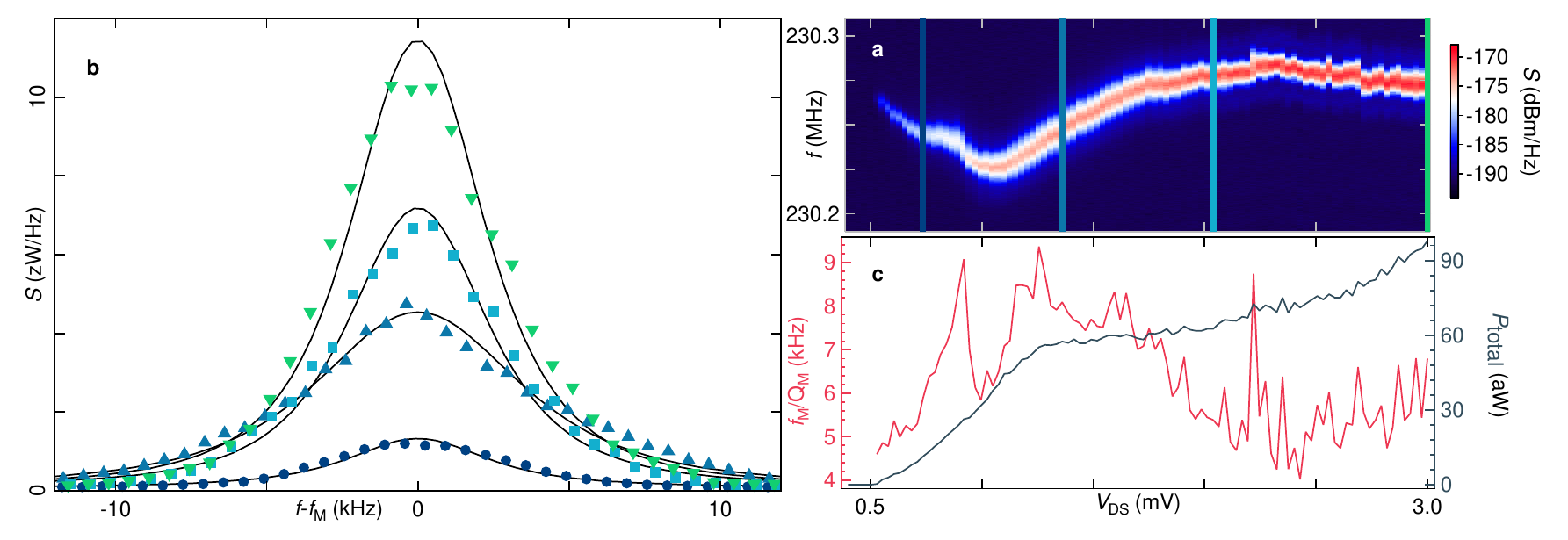}
\caption{\footnotesize{{\bf Mechanical response as a function of bias voltage}
{\bf a} Emission spectral density $S$ as a function of DC bias and frequency, with no RF excitation.
The gate voltage is fixed at $\VG = -1885.8$~mV.
{\bf b}, Points: Four selected spectra from {\bf a} (bias settings marked by vertical lines).
Curves: Lorentzian fits, with peak power and linewidth as free parameters.
{\bf c}, Peak power and linewidth $\fM/\QM$ across the entire range of {\bf a}, extracted by Lorentzian fits as in {\bf b}.
 }}
\label{FigS:linewidth}
\vspace{-0.5cm}
\end{figure*}

Since the oscillator is driven by the DC bias, increasing bias should lead to increasing emission power.
In Fig.~\ref{FigS:linewidth}, we test this prediction by measuring the emission spectrum without RF driving, as $\VDS$ is increased.
The onset of oscillations is at $\VDS\approx 0.5$~mV, and becomes stronger as expected (Fig.~\ref{FigS:linewidth}a).
By fitting the spectrum with a Lorentzian lineshape (Fig.~\ref{FigS:linewidth}b), we extract emission power and linewidth as a function of bias (Fig.~\ref{FigS:linewidth}c).
There is a slight line narrowing at high power, possibly because statistical fluctuations of the current become less significant.
As seen from the figure, the oscillator frequency varies slightly with bias, probably because the changing quantum dot charge modifies the nanotube's tension.

\subsection{Further characterization of the emission sidebands}
\begin{figure*}[ptb]
\includegraphics[width=\wwidth]{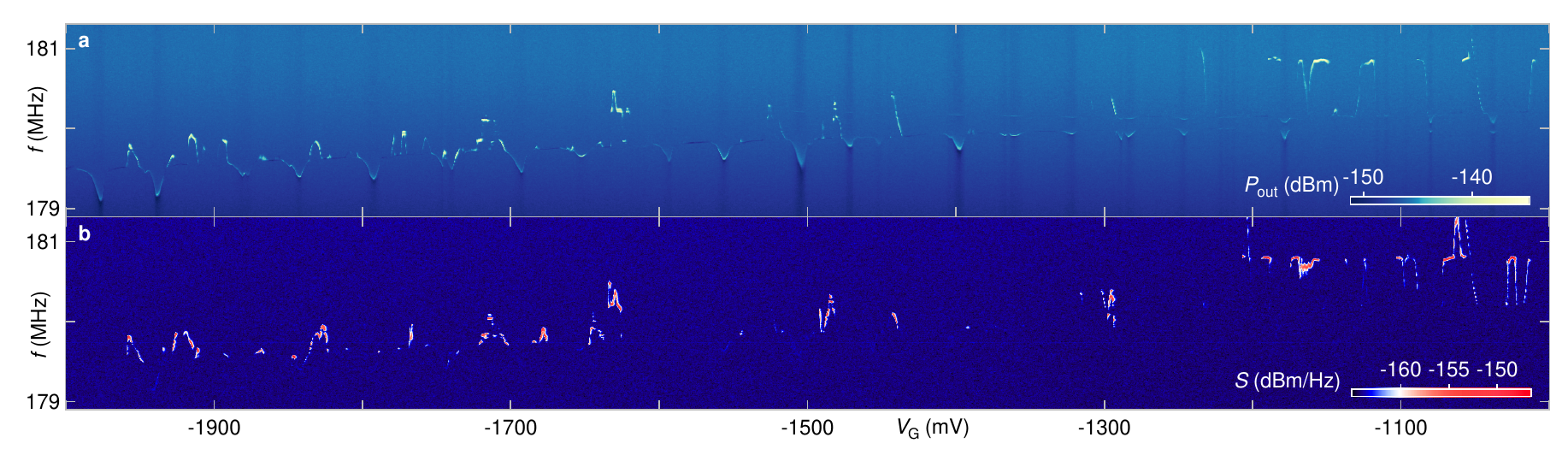}
\caption{\footnotesize{{\bf Mechanical resonance and oscillation in a previous cool-down.}
{\bf a}, Transmission measurement showing resonance, similar to Fig.~2b of the main text.
{\bf b}, Emission measurement showing oscillation, similar to Fig.~2c of the main text.
}}
\label{FigS:NoSQUIDSideband}
\vspace{-0.5cm}
\end{figure*}

In this section we discuss the origin of the various frequency sidebands appearing in the emission spectrum, and in particular the unexplained sidebands highlighted in Fig.~5a,\,b of the main text.
There are at least two expected effects that give rise to sidebands. Firstly, they can arise from mixing in the amplifier chain, since the main emission peak is then modulated by low-frequency electrical interference. These sidebands appear at fixed frequency offset from the main spectral peak. Secondly, they arise in an injection-locking experiment when the injection frequency is just outside the locking range~\cite{Siegman1986}. Although we observe both kinds of sideband, we now show that the highlighted sidebands in Fig.~5a,\,b do not fall into either category.

Sidebands due to mixing are seen in Fig.~2c of the main text, running parallel to the main emission peak.
These arise mainly from the SQUID amplifier, whose excellent sensitivity comes at the price of comparatively poor linearity.
We have confirmed this by (1) verifying that similar sidebands appear when an electrical tone is injected directly into the SQUID, bypassing the device altogether; and (2) measuring the nanotube oscillator on a separate cooldown with no SQUID mounted. The resulting transmission (Fig~\ref{FigS:NoSQUIDSideband}a) and emission (Fig~\ref{FigS:NoSQUIDSideband}b) show no visible sidebands.

A sideband due to injection locking is seen in Fig.~5b of the main text, running from from top left to bottom right of the panel outside the locked regime. This is a distortion sideband, as expected from the conventional theory of injection locking~\cite{Siegman1986}.

\begin{figure*}[pbt]
\includegraphics[width=\wwidth]{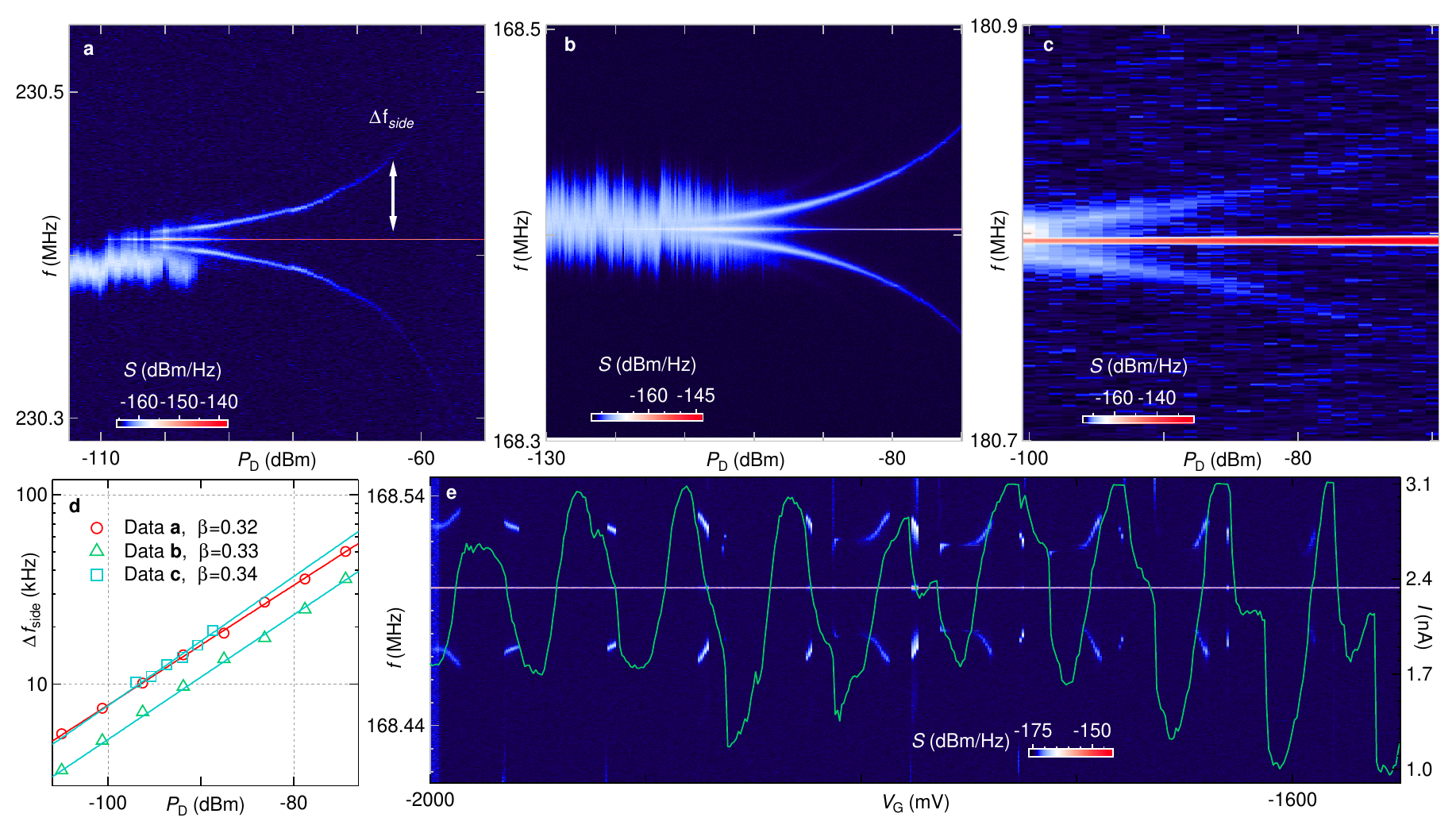}
\caption{\footnotesize{{\bf Sideband characterization.}
{\bf a,\,b,\,c}, Transmission spectrum plotted against injection power under different operating conditions.
In each plot, the nanotube is free-running at low power and locked at high power.
In the locked regime, the expected central emission peak is accompanied by an unexpected pari of sidebands.
Data in {\bf a} and {\bf b} were measured during the same cooldown as the main text, but with the SQUID unbiased; data in {\bf c} were measured in a previous cooldown in a separate cryostat with no SQUID fitted, and using a different RF source and spectrum analyser.
{\bf d}, Symbols: sideband spacing taken from panels {\bf a-c}.
Lines: fits to $\fside \propto \PD^\beta$, with $\beta$ as a free parameter.
{\bf e}, Colour plot: emission spectrum plotted against frequency (left axis) for different gate settings, with injection tone fixed in power and frequency.
Line plot: DC current (right axis) measured over the same gate range.
(The signal is clipped near $3$~nA due to current amplifier saturation.)
This figure shows that the sideband spacing $\fside$ is correlated with the current.
 }}
\label{FigS:SideBandPeaks}
\vspace{-0.5cm}
\end{figure*}

However, both panels in Fig.~5a,\,b also show sidebands inside the locking range, as highlighted by arrows in Fig.~5b.
To further investigate these sidebands, we measured the emission spectrum as a function of injection power (Fig.~\ref{FigS:SideBandPeaks}a-c).
This measurement was repeated for two different device configurations, and also on a prior cooldown without the SQUID amplifier.
All three measurements show the same unexpected sidebands, with the sideband spacing $\fside$ depending on the injection power.
We find that all three datasets are will fitted by the empirical function $\fside \propto \PD^\beta$ with $\beta \approx 0.3$ (Fig.~\ref{FigS:SideBandPeaks}d).
To further confirm that the sidebands originate in the device rather than in the apparatus, Fig.~\ref{FigS:SideBandPeaks}e shows the spectrum as a function of gate voltage.
The sidebands appear and disappear, presumably depending whether the oscillator is running, and furthermore the sideband spacing fluctuates.
A larger sideband spacing correlates with a larger DC current.
These sidebands remain unexplained and imply that the device is not completely described by the conventional theory of injection locking.

\subsection{Joint histogram of feedback-locked oscillation}
\begin{figure}[hptb]
\includegraphics[width=89mm]{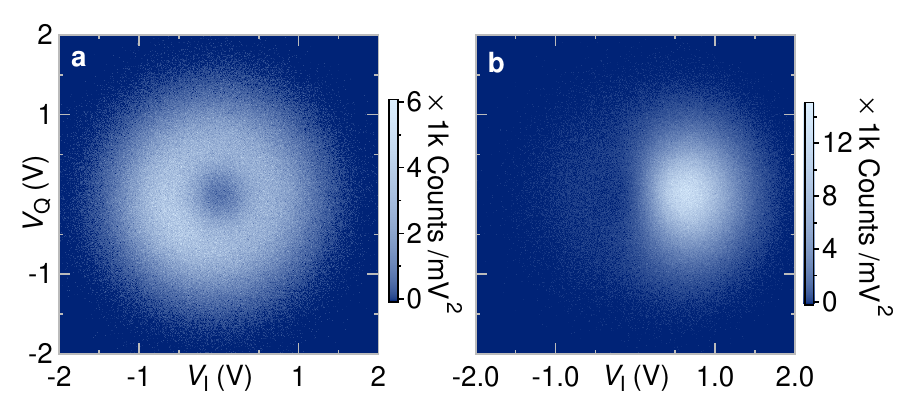}
\caption{\footnotesize{{\bf Effect of phase locking.}
{\bf a}, Joint histogram, similar to Fig.~3c, when the oscillator is above threshold and feedback is applied with unoptimized PID parameters: $P = 500$~kV/deg, $I = 500$~kV/(deg s), $D=0$.
The histogram is slightly asymmetric, with a higher weighting on the left.
{\bf b}, Similar data with optimized parameters: $P = 105$~kV/deg, $I = 200$~kV/(deg s), $D=0$.}}
\label{FigS:PIDHist}
\vspace{-0.5cm}
\end{figure}

Figure~\ref{FigS:PIDHist} shows the effect of feedback on the coherence histogram.
If the resonator were coherent but without a fixed phase, the joint histogram would show the same ring shape as Fig.~3c.
In a perfectly phase-locked-loop, the ring would collapse towards a single point on its perimeter.
Figure~\ref{FigS:PIDHist} shows this evolution.
When feedback is turned on without optimizing the PID parameters, the ring becomes slightly skewed (Fig.~\ref{FigS:PIDHist}{a}).
As the parameters are improved, the phase fluctuates less, and eventually the histogram collapses to the small arc of Fig.~\ref{FigS:PIDHist}{b}.

\bibliography{./ref}
\vfill

%% file: Cover.tex
\title{A coherent nanomechanical oscillator driven by single-electron tunnelling}

\author{Yutian~Wen}
\affiliation{Department of Materials, University of Oxford, Parks Road, Oxford OX1 3PH, United Kingdom}

\author{N.~Ares}
\affiliation{Department of Materials, University of Oxford, Parks Road, Oxford OX1 3PH, United Kingdom}

\author{F.J.~Schupp}
\affiliation{Department of Materials, University of Oxford, Parks Road, Oxford OX1 3PH, United Kingdom}

\author{T.~Pei}
\affiliation{Department of Materials, University of Oxford, Parks Road, Oxford OX1 3PH, United Kingdom}

\author{G.A.D.~Briggs}
\affiliation{Department of Materials, University of Oxford, Parks Road, Oxford OX1 3PH, United Kingdom}

\author{E.A.~Laird}
\email{e.a.laird@lancaster.ac.uk}
\affiliation{Department of Physics, Lancaster University, Lancaster, LA1 4YB, United Kingdom}
\affiliation{Department of Materials, University of Oxford, Parks Road, Oxford OX1 3PH, United Kingdom}

%% file: Definitions.tex
\newcommand{\VDS}{V_\text{DS}}
\newcommand{\VS}{V_\text{S}}
\newcommand{\VG}{V_\text{G}}
\newcommand{\VM}{V_\text{M}}
\newcommand{\Vout}{V_\text{out}}
\newcommand{\VN}{V_\text{N}}

\newcommand{\fE}{f_\text{E}}

\newcommand{\VI}{V_\text{I}}
\newcommand{\VQ}{V_\text{Q}}
\newcommand{\VT}{V_\text{T}}
\newcommand{\fbeat}{f_\text{beat}}

\newcommand{\IDS}{I}
\newcommand{\Pout}{\ensuremath{P_\mathrm{out} }}
\newcommand{\Sout}{S}
\newcommand{\RBW}{\Delta f}

\newcommand{\Ptot}{P_\text{total}}
\newcommand{\dfin}{\Delta f_\text{in}}

\newcommand{\fac}{f_\text{ac}}
\newcommand{\Vac}{V_\text{ac}}
\newcommand{\Voffset}{V_\text{offset}}

\newcommand{\deltaf}{\Delta f}
\newcommand{\ttherm}{T_\text{thermal}}
\newcommand{\deltafD}{\Delta f_\text{D}}
\newcommand{\gvv}{g^{(1)}}

\newcommand{\fside}{\Delta f_\text{side}}

\newcommand{\fM}{f_\text{M}}
\newcommand{\QM}{Q_\text{M}}
\newcommand{\fLO}{f_\text{LO}}

\newcommand{\CS}{C_\text{S}}
\newcommand{\CD}{C_\text{D}}
\newcommand{\CNT}{C_\text{NT}}
\newcommand{\Ztrans}{Z_\text{trans}}

\newcommand{\fD}{f_\text{D}}
\newcommand{\PD}{P_\text{D}}

\newcommand{\Fem}{F_\text{EM}}
\newcommand{\EQ}{\varepsilon_\text{Q}}
\newcommand{\uZPM}{u_\text{ZP}}
\newcommand{\FZPM}{F_\text{ZP}}
\newcommand{\sigtau}{\sigma_\text{Allan}(\tau)}
\newcommand{\fSA}{f_\text{SA}}

\newcommand{\Epolaron}{E_\text{p}}

\newcommand{\tauc}{\tau_\text{coh}}
\newcommand{\taufilter}{\tau_\text{filter}}

\newcommand*{\citen}[1]{%
  \begingroup
    \romannumeral-`\x 
    \setcitestyle{numbers}%
    \cite{#1}%
  \endgroup
}

\newcommand{\nphonon}{\bar{n}_\text{p}}
\newcommand{\fST}{f_\text{ST}}
\newcommand{\fWiseman}{f_\text{W}}
\newcommand{\iu}{\mathrm{i}\mkern1mu}
\newcommand{\Zzero}{Z_0}
\newcommand{\kB}{k_\text{B}}
\newcommand{\qprime}{q'}
\newcommand{\dIdVg}{\frac{\partial I}{\partial \VG}}
\newcommand{\uRMS}{u_\text{RMS}}
\newcommand{\OmegaZero}{\Omega_0}

%% file: ms.bbl
\begin{thebibliography}{10}
\expandafter\ifx\csname url\endcsname\relax
  \def\url#1{\texttt{#1}}\fi
\expandafter\ifx\csname urlprefix\endcsname\relax\def\urlprefix{URL }\fi
\providecommand{\bibinfo}[2]{#2}
\providecommand{\eprint}[2][]{\url{#2}}

\bibitem{Clerk2010}
\bibinfo{author}{Clerk, A.~A.} \emph{et~al.}
\newblock \bibinfo{title}{{Introduction to quantum noise, measurement, and
  amplification}}.
\newblock \emph{\bibinfo{journal}{Rev. Mod. Phys.}}
  \textbf{\bibinfo{volume}{82}}, \bibinfo{pages}{1155--1208}
  (\bibinfo{year}{2010}).

\bibitem{Schoelkopf1998a}
\bibinfo{author}{Schoelkopf, R.~J.}, \bibinfo{author}{Wahlgren, P.},
  \bibinfo{author}{Kozhevnikov, A.~A.}, \bibinfo{author}{Delsing, P.} \&
  \bibinfo{author}{Prober, D.~E.}
\newblock \bibinfo{title}{{The Radio-Frequency Single-Electron Transistor
  (RF-SET): A Fast and Ultrasensitive Electrometer}}.
\newblock \emph{\bibinfo{journal}{Science}} \textbf{\bibinfo{volume}{280}},
  \bibinfo{pages}{1238--1242} (\bibinfo{year}{1998}).

\bibitem{LaHaye2004}
\bibinfo{author}{LaHaye, M.~D.}, \bibinfo{author}{Buu, O.},
  \bibinfo{author}{Camarota, B.} \& \bibinfo{author}{Schwab, K.~C.}
\newblock \bibinfo{title}{{Approaching the Quantum Limit of a Nanomechanical
  Resonator}}.
\newblock \emph{\bibinfo{journal}{Science}} \textbf{\bibinfo{volume}{304}},
  \bibinfo{pages}{74--77} (\bibinfo{year}{2004}).

\bibitem{Mozyrsky2006}
\bibinfo{author}{Mozyrsky, D.}, \bibinfo{author}{Hastings, M.~B.} \&
  \bibinfo{author}{Martin, I.}
\newblock \bibinfo{title}{{Intermittent polaron dynamics: Born-Oppenheimer
  approximation out of equilibrium}}.
\newblock \emph{\bibinfo{journal}{Phys. Rev. B}} \textbf{\bibinfo{volume}{73}},
  \bibinfo{pages}{035104} (\bibinfo{year}{2006}).

\bibitem{Steele2009}
\bibinfo{author}{Steele, G.~A.} \emph{et~al.}
\newblock \bibinfo{title}{{Strong coupling between single-electron tunneling
  and nanomechanical motion.}}
\newblock \emph{\bibinfo{journal}{Science}} \textbf{\bibinfo{volume}{325}},
  \bibinfo{pages}{1103--1107} (\bibinfo{year}{2009}).

\bibitem{Lassagne2009}
\bibinfo{author}{Lassagne, B.}, \bibinfo{author}{Tarakanov, Y.},
  \bibinfo{author}{Kinaret, J.}, \bibinfo{author}{Daniel, G.~S.} \&
  \bibinfo{author}{Bachtold, A.}
\newblock \bibinfo{title}{{Coupling mechanics to charge transport in carbon
  nanotube mechanical resonators}}.
\newblock \emph{\bibinfo{journal}{Science}} \textbf{\bibinfo{volume}{325}},
  \bibinfo{pages}{1107--1110} (\bibinfo{year}{2009}).

\bibitem{Naik2006}
\bibinfo{author}{Naik, A.} \emph{et~al.}
\newblock \bibinfo{title}{{Cooling a nanomechanical resonator with quantum
  back-action}}.
\newblock \emph{\bibinfo{journal}{Nature}} \textbf{\bibinfo{volume}{443}},
  \bibinfo{pages}{193--196} (\bibinfo{year}{2006}).

\bibitem{Vahala2009}
\bibinfo{author}{Vahala, K.} \emph{et~al.}
\newblock \bibinfo{title}{{A phonon laser}}.
\newblock \emph{\bibinfo{journal}{Nat. Phys.}} \textbf{\bibinfo{volume}{5}},
  \bibinfo{pages}{682--686} (\bibinfo{year}{2009}).

\bibitem{Grudinin2010}
\bibinfo{author}{Grudinin, I.~S.}, \bibinfo{author}{Lee, H.},
  \bibinfo{author}{Painter, O.} \& \bibinfo{author}{Vahala, K.~J.}
\newblock \bibinfo{title}{{Phonon laser action in a tunable two-level system}}.
\newblock \emph{\bibinfo{journal}{Phys. Rev. Lett.}}
  \textbf{\bibinfo{volume}{104}}, \bibinfo{pages}{083901}
  (\bibinfo{year}{2010}).

\bibitem{Mahboob2013}
\bibinfo{author}{Mahboob, I.}, \bibinfo{author}{Nishiguchi, K.},
  \bibinfo{author}{Fujiwara, A.} \& \bibinfo{author}{Yamaguchi, H.}
\newblock \bibinfo{title}{{Phonon lasing in an electromechanical resonator}}.
\newblock \emph{\bibinfo{journal}{Phys. Rev. Lett.}}
  \textbf{\bibinfo{volume}{110}}, \bibinfo{pages}{127202}
  (\bibinfo{year}{2013}).

\bibitem{Sazonova2004}
\bibinfo{author}{Sazonova, V.} \emph{et~al.}
\newblock \bibinfo{title}{{A tunable carbon nanotube electromechanical
  oscillator}}.
\newblock \emph{\bibinfo{journal}{Nature}} \textbf{\bibinfo{volume}{431}},
  \bibinfo{pages}{284--287} (\bibinfo{year}{2004}).

\bibitem{Wen2018}
\bibinfo{author}{Wen, Y.}, \bibinfo{author}{Ares, N.}, \bibinfo{author}{Pei,
  T.}, \bibinfo{author}{Briggs, G. A.~D.} \& \bibinfo{author}{Laird, E.~A.}
\newblock \bibinfo{title}{{Measuring carbon nanotube vibrations using a
  single-electron transistor as a fast linear amplifier}}.
\newblock \emph{\bibinfo{journal}{Appl. Phys. Lett.}}
  \textbf{\bibinfo{volume}{113}}, \bibinfo{pages}{153101}
  (\bibinfo{year}{2018}).

\bibitem{DeBonis2018}
\bibinfo{author}{de~Bonis, S.~L.} \emph{et~al.}
\newblock \bibinfo{title}{{Ultrasensitive Displacement Noise Measurement of
  Carbon Nanotube Mechanical Resonators}}.
\newblock \emph{\bibinfo{journal}{Nano Letters}} \textbf{\bibinfo{volume}{18}},
  \bibinfo{pages}{5324} (\bibinfo{year}{2018}).

\bibitem{Khivrich2019}
\bibinfo{author}{Khivrich, I.}, \bibinfo{author}{Clerk, A.~A.} \&
  \bibinfo{author}{Ilani, S.}
\newblock \bibinfo{title}{{Nanomechanical pump–probe measurements of
  insulating electronic states in a carbon nanotube}}.
\newblock \emph{\bibinfo{journal}{Nat. Nanotechnol.}}
  \textbf{\bibinfo{volume}{14}}, \bibinfo{pages}{161--167}
  (\bibinfo{year}{2019}).

\bibitem{Armour2004}
\bibinfo{author}{Armour, A.~D.}, \bibinfo{author}{Blencowe, M.~P.} \&
  \bibinfo{author}{Zhang, Y.}
\newblock \bibinfo{title}{{Classical dynamics of a nanomechanical resonator
  coupled to a single-electron transistor}}.
\newblock \emph{\bibinfo{journal}{Phys. Rev. B}} \textbf{\bibinfo{volume}{69}}
  (\bibinfo{year}{2004}).

\bibitem{Rodrigues2007}
\bibinfo{author}{Rodrigues, D.~A.}, \bibinfo{author}{Imbers, J.} \&
  \bibinfo{author}{Armour, A.~D.}
\newblock \bibinfo{title}{{Quantum dynamics of a resonator driven by a
  superconducting single-electron transistor: A solid-state analogue of the
  micromaser}}.
\newblock \emph{\bibinfo{journal}{Physical Review Letters}}
  \textbf{\bibinfo{volume}{98}}, \bibinfo{pages}{067204}
  (\bibinfo{year}{2007}).

\bibitem{Bennett2006}
\bibinfo{author}{Bennett, S.~D.} \& \bibinfo{author}{Clerk, A.~A.}
\newblock \bibinfo{title}{{Laser-like instabilities in quantum
  nano-electromechanical systems}}.
\newblock \emph{\bibinfo{journal}{Phys. Rev. B}} \textbf{\bibinfo{volume}{74}},
  \bibinfo{pages}{201301} (\bibinfo{year}{2006}).

\bibitem{Usmani2007}
\bibinfo{author}{Usmani, O.}, \bibinfo{author}{Blanter, Y.~M.} \&
  \bibinfo{author}{Nazarov, Y.~V.}
\newblock \bibinfo{title}{{Strong feedback and current noise in
  nanoelectromechanical systems}}.
\newblock \emph{\bibinfo{journal}{Phys. Rev. B}} \textbf{\bibinfo{volume}{75}},
  \bibinfo{pages}{195312} (\bibinfo{year}{2007}).

\bibitem{Huttel2009}
\bibinfo{author}{H{\"{u}}ttel, A.~K.}, \bibinfo{author}{Witkamp, B.},
  \bibinfo{author}{Leijnse, M.}, \bibinfo{author}{Wegewijs, M.} \&
  \bibinfo{author}{{van der Zant}, H. S.~J.}
\newblock \bibinfo{title}{{Pumping of Vibrational Excitations in the
  Coulomb-Blockade Regime in a Suspended Carbon Nanotube}}.
\newblock \emph{\bibinfo{journal}{Phys. Rev. Lett.}}
  \textbf{\bibinfo{volume}{102}}, \bibinfo{pages}{225501}
  (\bibinfo{year}{2009}).

\bibitem{Eichler2011}
\bibinfo{author}{Eichler, A.}, \bibinfo{author}{Chaste, J.},
  \bibinfo{author}{Moser, J.} \& \bibinfo{author}{Bachtold, A.}
\newblock \bibinfo{title}{{Parametric Amplification and Self-Oscillation in a
  Nanotube Mechanical Resonator}}.
\newblock \emph{\bibinfo{journal}{Nano Letters}} \textbf{\bibinfo{volume}{11}},
  \bibinfo{pages}{2699--2703} (\bibinfo{year}{2011}).

\bibitem{Tsioutsios2017}
\bibinfo{author}{Tsioutsios, I.}, \bibinfo{author}{Tavernarakis, A.},
  \bibinfo{author}{Osmond, J.}, \bibinfo{author}{Verlot, P.} \&
  \bibinfo{author}{Bachtold, A.}
\newblock \bibinfo{title}{{Real-Time Measurement of Nanotube Resonator
  Fluctuations in an Electron Microscope}}.
\newblock \emph{\bibinfo{journal}{Nano Lett.}} \textbf{\bibinfo{volume}{17}},
  \bibinfo{pages}{1748--1755} (\bibinfo{year}{2017}).

\bibitem{Barnard2019}
\bibinfo{author}{Barnard, A.~W.}, \bibinfo{author}{Zhang, M.},
  \bibinfo{author}{Wiederhecker, G.~S.}, \bibinfo{author}{Lipson, M.} \&
  \bibinfo{author}{McEuen, P.~L.}
\newblock \bibinfo{title}{{Real-time vibrations of a carbon nanotube}}.
\newblock \emph{\bibinfo{journal}{Nature}}  (\bibinfo{year}{2019}).

\bibitem{Wu2010}
\bibinfo{author}{Wu, C.~C.}, \bibinfo{author}{Liu, C.~H.} \&
  \bibinfo{author}{Zhong, Z.}
\newblock \bibinfo{title}{{One-step direct transfer of pristine single-walled
  Carbon nanotubes for functional nanoelectronics}}.
\newblock \emph{\bibinfo{journal}{Nano Lett.}} \textbf{\bibinfo{volume}{10}},
  \bibinfo{pages}{1032--1036} (\bibinfo{year}{2010}).

\bibitem{Schupp2018}
\bibinfo{author}{Schupp, F.~J.} \emph{et~al.}
\newblock \bibinfo{title}{{Radio-frequency reflectometry of a quantum dot using
  an ultra-low-noise SQUID amplifier}}  (\bibinfo{year}{2018}).
\newblock \eprint{Preprint at https://arxiv.org/abs/1810.05767}.

\bibitem{FoxBook}
\bibinfo{author}{Fox, M.}
\newblock \emph{\bibinfo{title}{Quantum Optics: An Introduction}}
  (\bibinfo{publisher}{Oxford University Press}, \bibinfo{year}{2006}).

\bibitem{Liu2015Science}
\bibinfo{author}{Liu, Y.-Y.} \emph{et~al.}
\newblock \bibinfo{title}{{Semiconductor double quantum dot micromaser}}.
\newblock \emph{\bibinfo{journal}{Science}} \textbf{\bibinfo{volume}{347}},
  \bibinfo{pages}{285--287} (\bibinfo{year}{2015}).

\bibitem{Cassidy}
\bibinfo{author}{Cassidy, M.~C.} \emph{et~al.}
\newblock \bibinfo{title}{{Demonstration of an ac Josephson junction laser}}.
\newblock \emph{\bibinfo{journal}{Science}} \textbf{\bibinfo{volume}{355}},
  \bibinfo{pages}{939--942} (\bibinfo{year}{2017}).

\bibitem{Pistolesi}
\bibinfo{author}{Pistolesi, F.}, \bibinfo{author}{Blanter, Y.~M.} \&
  \bibinfo{author}{Martin, I.}
\newblock \bibinfo{title}{{Self-consistent theory of molecular switching}}.
\newblock \emph{\bibinfo{journal}{Phys. Rev. B}} \textbf{\bibinfo{volume}{78}},
  \bibinfo{pages}{085127} (\bibinfo{year}{2008}).

\bibitem{Adler1946}
\bibinfo{author}{Adler, R.}
\newblock \bibinfo{title}{{A Study of Locking Phenomena in Oscillators}}.
\newblock \emph{\bibinfo{journal}{Proc. IRE}} \textbf{\bibinfo{volume}{34}},
  \bibinfo{pages}{351--357} (\bibinfo{year}{1946}).

\bibitem{Stover1966}
\bibinfo{author}{Stover, H.~L.} \& \bibinfo{author}{Steier, W.~H.}
\newblock \bibinfo{title}{{Locking of Laser Oscillators by Light Injection}}.
\newblock \emph{\bibinfo{journal}{Appl. Phys. Lett.}}
  \textbf{\bibinfo{volume}{8}}, \bibinfo{pages}{91} (\bibinfo{year}{1966}).

\bibitem{Liu2015PRA}
\bibinfo{author}{Liu, Y.-Y.}, \bibinfo{author}{Stehlik, J.},
  \bibinfo{author}{Gullans, M.~J.}, \bibinfo{author}{Taylor, J.~M.} \&
  \bibinfo{author}{Petta, J.~R.}
\newblock \bibinfo{title}{{Injection locking of a semiconductor
  double-quantum-dot micromaser}}.
\newblock \emph{\bibinfo{journal}{Phys. Rev. A}} \textbf{\bibinfo{volume}{92}},
  \bibinfo{pages}{053802} (\bibinfo{year}{2015}).

\bibitem{Knunz2010}
\bibinfo{author}{Kn{\"{u}}nz, S.} \emph{et~al.}
\newblock \bibinfo{title}{{Injection locking of a trapped-ion phonon laser}}.
\newblock \emph{\bibinfo{journal}{Phys. Rev. Lett.}}
  \textbf{\bibinfo{volume}{105}}, \bibinfo{pages}{013004}
  (\bibinfo{year}{2010}).

\bibitem{Seitner2017}
\bibinfo{author}{Seitner, M.~J.}, \bibinfo{author}{Abdi, M.},
  \bibinfo{author}{Ridolfo, A.}, \bibinfo{author}{Hartmann, M.~J.} \&
  \bibinfo{author}{Weig, E.~M.}
\newblock \bibinfo{title}{{Parametric Oscillation, Frequency Mixing, and
  Injection Locking of Strongly Coupled Nanomechanical Resonator Modes}}.
\newblock \emph{\bibinfo{journal}{Physical Review Letters}}
  \textbf{\bibinfo{volume}{118}}, \bibinfo{pages}{254301}
  (\bibinfo{year}{2017}).

\bibitem{Drever1983}
\bibinfo{author}{Drever, R.~W.} \emph{et~al.}
\newblock \bibinfo{title}{{Laser phase and frequency stabilization using an
  optical resonator}}.
\newblock \emph{\bibinfo{journal}{Appl. Phys. B}}
  \textbf{\bibinfo{volume}{31}}, \bibinfo{pages}{97--105}
  (\bibinfo{year}{1983}).

\bibitem{Abbott2009}
\bibinfo{author}{{Abbott, B.\ P. \emph{et al.}}}
\newblock \bibinfo{title}{{LIGO : the Laser Interferometer Gravitational-Wave
  Observatory}}.
\newblock \emph{\bibinfo{journal}{Rep. Prog. Phys.}}
  \textbf{\bibinfo{volume}{72}}, \bibinfo{pages}{076901}
  (\bibinfo{year}{2009}).

\bibitem{Schawlow1958}
\bibinfo{author}{Schawlow, A.~L.} \& \bibinfo{author}{Townes, C.~H.}
\newblock \bibinfo{title}{{Infrared and optical masers}}.
\newblock \emph{\bibinfo{journal}{Phys. Rev.}} \textbf{\bibinfo{volume}{112}},
  \bibinfo{pages}{1940} (\bibinfo{year}{1958}).

\bibitem{Wiseman1999}
\bibinfo{author}{Wiseman, H.~M.}
\newblock \bibinfo{title}{{Light amplification without stimulated emission:
  Beyond the standard quantum limit to the laser linewidth}}.
\newblock \emph{\bibinfo{journal}{Phys. Rev. A}} \textbf{\bibinfo{volume}{60}},
  \bibinfo{pages}{4083} (\bibinfo{year}{1999}).

\bibitem{Cundiff2003}
\bibinfo{author}{Cundiff, S.~T.} \& \bibinfo{author}{Ye, J.}
\newblock \bibinfo{title}{{Colloquium: Femtosecond optical frequency combs}}.
\newblock \emph{\bibinfo{journal}{Reviews of Modern Physics}}
  \textbf{\bibinfo{volume}{75}}, \bibinfo{pages}{325--342}
  (\bibinfo{year}{2003}).

\bibitem{Erickson2014}
\bibinfo{author}{Erickson, R.~P.}, \bibinfo{author}{Vissers, M.~R.},
  \bibinfo{author}{Sandberg, M.}, \bibinfo{author}{Jefferts, S.~R.} \&
  \bibinfo{author}{Pappas, D.~P.}
\newblock \bibinfo{title}{{Frequency comb generation in superconducting
  resonators}}.
\newblock \emph{\bibinfo{journal}{Phys. Rev. Lett.}}
  \textbf{\bibinfo{volume}{113}}, \bibinfo{pages}{187002}
  (\bibinfo{year}{2014}).

\bibitem{Solinas2015}
\bibinfo{author}{Solinas, P.}, \bibinfo{author}{Gasparinetti, S.},
  \bibinfo{author}{Golubev, D.} \& \bibinfo{author}{Giazotto, F.}
\newblock \bibinfo{title}{{A Josephson radiation comb generator}}.
\newblock \emph{\bibinfo{journal}{Scientific Reports}}
  \textbf{\bibinfo{volume}{5}}, \bibinfo{pages}{12260} (\bibinfo{year}{2015}).

\bibitem{Chaste2012}
\bibinfo{author}{Chaste, J.} \emph{et~al.}
\newblock \bibinfo{title}{{A nanomechanical mass sensor with yoctogram
  resolution}}.
\newblock \emph{\bibinfo{journal}{Nat. Nanotechnol.}}
  \textbf{\bibinfo{volume}{7}}, \bibinfo{pages}{301--304}
  (\bibinfo{year}{2012}).

\bibitem{Stipe2001}
\bibinfo{author}{Stipe, B.~C.} \emph{et~al.}
\newblock \bibinfo{title}{{Electron Spin Relaxation Near a Micron-Size
  Ferromagnet}}.
\newblock \emph{\bibinfo{journal}{Phys. Rev. Lett.}}
  \textbf{\bibinfo{volume}{87}}, \bibinfo{pages}{277602}
  (\bibinfo{year}{2001}).

\bibitem{Maryam2013}
\bibinfo{author}{Maryam, W.}, \bibinfo{author}{Akimov, A.~V.},
  \bibinfo{author}{Campion, R.~P.} \& \bibinfo{author}{Kent, A.~J.}
\newblock \bibinfo{title}{{Dynamics of a vertical cavity quantum cascade phonon
  laser structure}}.
\newblock \emph{\bibinfo{journal}{Nature Communications}}
  \textbf{\bibinfo{volume}{4}}, \bibinfo{pages}{2184} (\bibinfo{year}{2013}).

\bibitem{Wiseman1997}
\bibinfo{author}{Wiseman, H.~M.}
\newblock \bibinfo{title}{{Defining the (atom) laser}}.
\newblock \emph{\bibinfo{journal}{Phys. Rev. A}} \textbf{\bibinfo{volume}{56}},
  \bibinfo{pages}{2068} (\bibinfo{year}{1997}).

\bibitem{Ottl2005}
\bibinfo{author}{\"{O}ttl, A.}, \bibinfo{author}{Ritter, S.},
  \bibinfo{author}{K{\"{o}}hl, M.} \& \bibinfo{author}{Esslinger, T.}
\newblock \bibinfo{title}{{Correlations and counting statistics of an atom
  laser}}.
\newblock \emph{\bibinfo{journal}{Phys. Rev. Lett.}}
  \textbf{\bibinfo{volume}{95}}, \bibinfo{pages}{090404}
  (\bibinfo{year}{2005}).

\bibitem{Brandes2003}
\bibinfo{author}{Brandes, T.} \& \bibinfo{author}{Lambert, N.}
\newblock \bibinfo{title}{{Steering of a bosonic mode with a double quantum
  dot}}.
\newblock \emph{\bibinfo{journal}{Phys. Rev. B}} \textbf{\bibinfo{volume}{67}},
  \bibinfo{pages}{125323} (\bibinfo{year}{2003}).

\bibitem{Ohm2012}
\bibinfo{author}{Ohm, C.}, \bibinfo{author}{Stampfer, C.},
  \bibinfo{author}{Splettstoesser, J.} \& \bibinfo{author}{Wegewijs, M.}
\newblock \bibinfo{title}{{Readout of carbon nanotube vibrations based on
  spin-phonon coupling}}.
\newblock \emph{\bibinfo{journal}{Appl. Phys. Lett.}}
  \textbf{\bibinfo{volume}{100}}, \bibinfo{pages}{143103}
  (\bibinfo{year}{2012}).

\bibitem{Palyi2012}
\bibinfo{author}{P{\'{a}}lyi, A.}, \bibinfo{author}{Struck, P.~R.},
  \bibinfo{author}{Rudner, M.~S.}, \bibinfo{author}{Flensberg, K.} \&
  \bibinfo{author}{Burkard, G.}
\newblock \bibinfo{title}{{Spin-Orbit-Induced Strong Coupling of a Single Spin
  to a Nanomechanical Resonator}}.
\newblock \emph{\bibinfo{journal}{Phys. Rev. Lett.}}
  \textbf{\bibinfo{volume}{108}}, \bibinfo{pages}{206811}
  (\bibinfo{year}{2012}).

\bibitem{Ogata1970}
\bibinfo{author}{Ogata, K.}
\newblock \emph{\bibinfo{title}{{Modern control engineering}}}
  (\bibinfo{publisher}{Prentice Hall}, \bibinfo{year}{1970}).

\end{thebibliography}


\begin{thebibliography}{10}
\expandafter\ifx\csname url\endcsname\relax
  \def\url#1{\texttt{#1}}\fi
\expandafter\ifx\csname urlprefix\endcsname\relax\def\urlprefix{URL }\fi
\providecommand{\bibinfo}[2]{#2}
\providecommand{\eprint}[2][]{\url{#2}}

\bibitem{Bennett2006}
\bibinfo{author}{Bennett, S.~D.} \& \bibinfo{author}{Clerk, A.~A.}
\newblock \bibinfo{title}{{Laser-like instabilities in quantum
  nano-electromechanical systems}}.
\newblock \emph{\bibinfo{journal}{Phys. Rev. B}} \textbf{\bibinfo{volume}{74}},
  \bibinfo{pages}{201301} (\bibinfo{year}{2006}).

\bibitem{Usmani2007}
\bibinfo{author}{Usmani, O.}, \bibinfo{author}{Blanter, Y.~M.} \&
  \bibinfo{author}{Nazarov, Y.~V.}
\newblock \bibinfo{title}{{Strong feedback and current noise in
  nanoelectromechanical systems}}.
\newblock \emph{\bibinfo{journal}{Phys. Rev. B}} \textbf{\bibinfo{volume}{75}},
  \bibinfo{pages}{195312} (\bibinfo{year}{2007}).

\bibitem{Blanter2004}
\bibinfo{author}{Blanter, Y.~M.}, \bibinfo{author}{Usmani, O.} \&
  \bibinfo{author}{Nazaroy, Y.~V.}
\newblock \bibinfo{title}{{Erratum: Single-electron tunneling with strong
  mechanical feedback (Phys. Rev. Lett. (2004) 93 (136802))}}.
\newblock \emph{\bibinfo{journal}{Phys. Rev. Lett.}}
  \textbf{\bibinfo{volume}{94}}, \bibinfo{pages}{136802}
  (\bibinfo{year}{2005}).

\bibitem{Steele2009}
\bibinfo{author}{Steele, G.~A.} \emph{et~al.}
\newblock \bibinfo{title}{{Strong coupling between single-electron tunneling
  and nanomechanical motion.}}
\newblock \emph{\bibinfo{journal}{Science}} \textbf{\bibinfo{volume}{325}},
  \bibinfo{pages}{1103--1107} (\bibinfo{year}{2009}).

\bibitem{Lassagne2009}
\bibinfo{author}{Lassagne, B.}, \bibinfo{author}{Tarakanov, Y.},
  \bibinfo{author}{Kinaret, J.}, \bibinfo{author}{Daniel, G.~S.} \&
  \bibinfo{author}{Bachtold, A.}
\newblock \bibinfo{title}{{Coupling mechanics to charge transport in carbon
  nanotube mechanical resonators}}.
\newblock \emph{\bibinfo{journal}{Science}} \textbf{\bibinfo{volume}{325}},
  \bibinfo{pages}{1107--1110} (\bibinfo{year}{2009}).

\bibitem{Meerwaldt2012}
\bibinfo{author}{Meerwaldt, H.~B.} \emph{et~al.}
\newblock \bibinfo{title}{{Probing the charge of a quantum dot with a
  nanomechanical resonator}}.
\newblock \emph{\bibinfo{journal}{Phys. Rev. B}} \textbf{\bibinfo{volume}{86}},
  \bibinfo{pages}{115454} (\bibinfo{year}{2012}).

\bibitem{Barton2012}
\bibinfo{author}{Barton, R.~A.} \emph{et~al.}
\newblock \bibinfo{title}{{Photothermal self-oscillation and laser cooling of
  graphene optomechanical systems}}.
\newblock \emph{\bibinfo{journal}{Nano Letters}} \textbf{\bibinfo{volume}{12}},
  \bibinfo{pages}{4681--4686} (\bibinfo{year}{2012}).

\bibitem{Tavernarakis2018}
\bibinfo{author}{Tavernarakis, A.} \emph{et~al.}
\newblock \bibinfo{title}{{Optomechanics with a hybrid carbon nanotube
  resonator}}.
\newblock \emph{\bibinfo{journal}{Nat. Commun.}} \textbf{\bibinfo{volume}{9}},
  \bibinfo{pages}{1--8} (\bibinfo{year}{2018}).

\bibitem{Wen2018}
\bibinfo{author}{Wen, Y.}, \bibinfo{author}{Ares, N.}, \bibinfo{author}{Pei,
  T.}, \bibinfo{author}{Briggs, G. A.~D.} \& \bibinfo{author}{Laird, E.~A.}
\newblock \bibinfo{title}{{Measuring carbon nanotube vibrations using a
  single-electron transistor as a fast linear amplifier}}.
\newblock \emph{\bibinfo{journal}{Appl. Phys. Lett.}}
  \textbf{\bibinfo{volume}{113}}, \bibinfo{pages}{153101}
  (\bibinfo{year}{2018}).

\bibitem{Micchi2015}
\bibinfo{author}{Micchi, G.}, \bibinfo{author}{Avriller, R.} \&
  \bibinfo{author}{Pistolesi, F.}
\newblock \bibinfo{title}{{Mechanical Signatures of the Current Blockade
  Instability in Suspended Carbon Nanotubes}}.
\newblock \emph{\bibinfo{journal}{Phys. Rev. Lett.}}
  \textbf{\bibinfo{volume}{115}}, \bibinfo{pages}{206802}
  (\bibinfo{year}{2015}).

\bibitem{Schawlow1958}
\bibinfo{author}{Schawlow, A.~L.} \& \bibinfo{author}{Townes, C.~H.}
\newblock \bibinfo{title}{{Infrared and optical masers}}.
\newblock \emph{\bibinfo{journal}{Phys. Rev.}} \textbf{\bibinfo{volume}{112}},
  \bibinfo{pages}{1940} (\bibinfo{year}{1958}).

\bibitem{Wiseman1999}
\bibinfo{author}{Wiseman, H.~M.}
\newblock \bibinfo{title}{{Light amplification without stimulated emission:
  Beyond the standard quantum limit to the laser linewidth}}.
\newblock \emph{\bibinfo{journal}{Phys. Rev. A}} \textbf{\bibinfo{volume}{60}},
  \bibinfo{pages}{4083} (\bibinfo{year}{1999}).

\bibitem{Schupp2018}
\bibinfo{author}{Schupp, F.~J.} \emph{et~al.}
\newblock \bibinfo{title}{{Radio-frequency reflectometry of a quantum dot using
  an ultra-low-noise SQUID amplifier}}  (\bibinfo{year}{2018}).
\newblock \eprint{Preprint at https://arxiv.org/abs/1810.05767}.

\bibitem{Siegman1986}
\bibinfo{author}{{Siegman, A. E.}}
\newblock \emph{\bibinfo{title}{{Lasers}}} (\bibinfo{publisher}{University
  Science Books}, \bibinfo{year}{1986}).

\end{thebibliography}
